\documentclass[letterpaper,twocolumn,10pt]{article}
\usepackage{usenix2019_v3}
\usepackage[T1]{fontenc}
\usepackage[scaled=0.75]{beramono}
\usepackage{amsmath,amsthm,amssymb}
\usepackage{mathptmx}
\usepackage[ruled,vlined]{algorithm2e}
\usepackage{listings}
\usepackage{xcolor}
\usepackage{colortbl}
\usepackage{balance}
\usepackage{graphicx}
\usepackage[numbers,sort]{natbib}
\usepackage{wrapfig}
\usepackage[font={footnotesize}]{caption}
\usepackage{subcaption}
\usepackage{array}
\usepackage{booktabs}
\usepackage[export]{adjustbox}
\usepackage{datetime}
\usepackage{enumitem}
\usepackage{multirow}
\usepackage[frozencache]{minted}
\usepackage[binary-units=true]{siunitx}
\usepackage{pifont}
\usepackage{tikz}
\usetikzlibrary{tikzmark,calc}

\graphicspath{{./}{figure/}}

\setminted{breaklines=true,breaksymbolleft=,fontsize=\footnotesize}
\DeclareMathAlphabet{\mathcal}{OMS}{cmsy}{m}{n}

\newcolumntype{R}[2]{%
  >{\adjustbox{angle=#1,lap=\width-(#2)}\bgroup}%
  l%
  <{\egroup}%
}

\DeclareSIUnit{\operation}{op}

\hypersetup{
  bookmarks,
  colorlinks,
  urlcolor=magenta,
  citecolor=blue,
  linkcolor=blue,
}

\begin{document}

\date{}

\title{\Large \bf Observations on Porting In-memory KV stores to Persistent Memory}

\author{
{\rm Brian Choi}\\
Johns Hopkins University
\and
{\rm Parv Saxena}\\
Johns Hopkins University
 \and
 {\rm Ryan Huang}\\
Johns Hopkins University
\and
{\rm Randal Burns}\\
Johns Hopkins University
} 

\maketitle

\begin{abstract}

Systems that require high-throughput and fault tolerance, such as key-value stores and databases,
are looking to persistent memory to combine the performance of in-memory systems with
the data-consistent fault-tolerance of non-volatile stores. Persistent memory devices
provide fast byte-addressable access to non-volatile memory.  


We analyze the design space when integrating persistent memory into in-memory key value 
stores and quantify performance tradeoffs between throughput, latency, and and recovery time. 
Previous works have explored many design choices, but did not quantify the tradeoffs.
We implement persistent memory support in Redis and Memcached, adapting the 
data structures of each to work in two modes: (1) with all data in persistent 
memory and (2) a hybrid mode that uses persistent memory for key/value data and 
non-volatile memory for indexing and metadata. 
Our experience reveals three actionable design principles that hold in Redis and Memcached, 
despite their very different implementations. We conclude that the hybrid design increases throughput
and decreases latency at a minor cost in recovery time and code complexity.

\end{abstract}

\section{Introduction}
Persistent memory (PM) has emerged as a new class of storage technology, filling 
the gap between DRAM and SSDs. PM devices can be placed alongside DRAM on the processor 
memory bus to enable byte-addressable memory accesses, with latency comparable 
but slower (2--3$\times$ for loads~\cite{izraelevitz2019basic}) than DRAM, and 
10--100$\times$ faster than state-of-the-art NAND flash~\cite{zssd}. Unlike 
volatile DRAM, data stored in persistent memory will survive reboot and power loss. 
With Intel Optane DC Persistent Memory, the first PM product, being released 
recently~\cite{optane_dc_announcement}, many developers are looking to take advantage 
of persistent memory when building their applications. For in-memory key-value stores 
like Redis, it adds more efficient large capacity deployments and persistence 
options~\cite{RedisPMEMBenefits}. For non-volatile stores like Cassandra, it offers 
improved performance when compared with SSDs~\cite{Cassandra}.


The many efforts to integrate persistent memory into existing storage systems
have lead to a confusing array of design alternatives.  Systems choose from
among moving a volatile in-memory system to persistent memory, replacing SSDs
with persistent memory in a storage hierarchy, using write-ahead logging
principles to coordinate volatile and non-volatile memory, or decomposing the
system into volatile and persistent data and data structures. All choices have
consequences on latency, recovery time, and system complexity. 
Table~\ref{tab:taxonomy_persistence} categorizes existing work based on 
how they use persistent memory. No clear guidelines exist that define and quantify 
the trade-offs accompanying different design choices. 

\begin{table*}[t]
\footnotesize
\centering
\begin{tabular}{| c | p{2.5cm} | p{3cm} | p{3cm}| p{3cm}| } 
 \hline
 & Fully Persistent & Checkpoint/Logging & Hybrid (Persistent Data/Volatile Indexing) & 
Mixed Indexing and Data \\ 
 \hline
 Newly Designed 
 & \cite{Zuo:2018:WHH:3291168.3291202,Debnath:2015:RHT:2819001.2819002,227812,chen2011rethinking,Chen:2015:PBT:2752939.2752947,Chi:2014:MBT:2627369.2627630} 
  & \cite{Huang:2018:CPG:3277355.3277448}
 & \cite{KourtisFAST2019,wu2016nvmcached} 
 & \cite{Kim:2017:PHP:3126908.3126943,Xia:2017:HHI:3154690.3154724,Oukid:2016:FHS:2882903.2915251,yang2015nv,bailey2013exploring} \\ 
 \hline
 Modification to Existing 
 & \cite{MaratheHotStorage2017,NalliASPLOS2017}
 & \cite{libpmemlog-AOF} 
 & \cite{intelpmemRedis,LenovoMemcached,SnalliRedis} 
 & \cite{PMEMRedis,RedisFlash}\\
 \hline
\end{tabular}%
\caption{Taxonomy of persistent memory KV stores. {\em Fully persistent}
  systems place data and indexes in persistent memory.  {\em Hybrid} systems
  place data in persistent memory and maintain indexes in volatile memory.
  {\em Mixed} maintains indexes and/or data using a combination of volatile and
  persistent memory.} 
\label{tab:taxonomy_persistence}
\end{table*}

In response, we provide an extensive measurement of different high-level 
system designs in order to inform developers and the research community about 
these trade-offs.
Our analysis focuses on designs that integrate persistent memory into existing 
in-memory key-value stores in order to provide data-consistent crash recovery. 
This approach is popular, because it is incremental, involves small code changes, and allows
high-throughput systems to add fault tolerance.
We examine three major design decisions that developers have to make when porting 
their systems to persistent memory: (1) \emph{What data structures should be persistent 
and what data structures should be volatile?} (2) \emph{What should 
the granularity of persistence be?} (3) \emph{What PM primitives and interfaces to use?} 
We then study how these design decisions directly effect six system properties, 
including operational throughput and latency, tail latency, 
recovery time, exposure to data loss, scalability and system complexity. 



We choose two popular in-memory key-value stores, Redis and Memcached, as our subject. 
As there are already multiple existing works~\cite{MaratheHotStorage2017,NalliASPLOS2017,PMEMRedis,
intelpmemRedis,LenovoMemcached}, our initial attempt was to leverage these implementations 
and compare them. However, these systems encode many decisions regarding one system property;
they are incomplete or incomparable to evaluate by other properties. For example, many implementations 
focus on the operational performance aspect and are not fully recoverable upon restart. 
We were also unable to gain the source code for several published works.

Therefore, we decide to implement two different and comparable high-level systems designs in 
both Redis and Memcached. The first design is {\em fully persistent} because all 
volatile data structures are maintained in persistent memory. This eliminates most 
recovery work at the cost of additional latency when writing to or extending the 
hash table. The second {\em hybrid} design places key-value data in persistent memory 
and keeps the hash table indexing structure in volatile memory.  During recovery, the 
hash table must be reconstructed from the keys and values. By implementing these four 
versions, we are able to isolate and compare the effect on individual design choice.
We show that our conclusions hold across these very different implementations. 

We summarize three actionable design principles for porting volatile 
key/value stores to persistent memory:
\begin{itemize} 
\addtolength{\itemsep}{-5pt}
\item A hybrid design nearly doubles operational throughput at the cost of an increased time
to recovery that varies by system from minor in Memcached to major in Redis. System designers
have a nuanced choice between operational performance and recovery. 
\item Allocating data in large chunks reduces latency by amortizing allocation costs and increases 
recovery performance. 
\item Full featured persistent memory libraries ease developement and lead to simple implementations.  
\end{itemize}

\section{Background}
PM technologies like 3D XPoint and ReRAM promise to provide byte-addressable accesses with 
low latency, unlike SSDs and disks that perform I/O at a block-granularity. After years 
of anticipation, the Intel Optane DC Persistent 
Memory product became publicly available in April 2019. Optane DC offers two 
operation modes~\cite{app_direct_mode}. The \emph{Memory Mode} transparently 
integrates the device into the memory hierarchy 
so that applications perceive the device as a large pool of main
memory. The advantage is that no application changes are required. But the data is 
not durable upon power loss. In the \emph{App Direct Mode}, applications are aware 
of the PM, and data written to the PM can be persisted. But applications 
need to be modified to access the persistent memory region via a PM-aware 
file system or loads/stores. 



It is expected that PM devices like Optane DC (particularly its App Direct Mode)
would not only enable a rising number of new PM storage systems but also motivate 
developers of existing (legacy) in-memory applications to modify their
applications to build efficient persistence. We focus on the latter ``porting'' 
scenario. A representative class of applications is existing in-memory key-value stores. 
At present, in-memory key-value stores support persistence through either periodic 
snapshots, which can lose significant data, or costly logging. 

Porting existing in-memory key-value stores to PM, however, has complexities that arise 
from hardware characteristics. First, PM has much higher write latency~\cite{Mnemosyne2011ASPLOS} 
and lower write bandwidth~\cite{izraelevitz2019basic} compared to DRAM.
It is thus not feasible to port all volatile data that involves frequent 
writes or updates to persistent memory. This suggests that developers 
have to make careful decisions to selectively persist data structures.
Second, PM requires the proper use of flushes, fences, and transactions
to ensure data consistency. Developers need to explicitly flush cache lines using 
instructions like {\tt clwb} because writes to the PM device may be cached. Besides 
flush, {\tt sfence} is necessary because the compiler or memory controller may reorder writes. 
While the PM device guarantees failure atomicity in 8-byte units, larger writes could 
lead to inconsistent data in the event of failure, which should be avoided with 
transactions. All of these mechanisms introduce significant performance 
overhead~\cite{Chen:2015:PBT:2752939.2752947, condit2009better,Venkataraman:2011:CDD:1960475.1960480} 
that must be considered when porting a key-value store.

\vspace{5pt}

\noindent{\bf Redis:}
Redis is one of the most popular key value stores \cite{db} and used as an
in-memory cache or as a database.  Redis stores data in main memory for fast
access and implements an extendable hash table indexing structure for efficient
lookups.  Redis extends the key/value model to support a large number of data
structures, such as string, hashmaps, sets, and lists. Notably, Redis is a
single threaded service, which eliminates the need for locks and
synchronization.  Redis supports the \emph{RDB} feature that takes periodic 
snapshots of the dataset either after a period of time or after a number of 
keys have been modified. It also provides the \emph{AOF} persistence option 
in which Redis logs every write operation and replays the log on startup. 

\vspace{5pt}

\noindent{\bf Memcached:}
Memcached is another popular key value store used primarily as a cache. Similar
to Redis, Memcached keeps key value data in memory indexed by an extendible
hash table. Memcached does not have built-in persistence, but there are
extensions.  
Memcached differs from Redis in several important ways.  First, it uses a slab
allocation organized by data size to amortize allocation across many objects.
Second, Memcached supports multithreading and uses locks to synchronize concurrent
access to data structures. Third, Memcached does not have complete persistence options like Redis does yet~\cite{restartable_memcached}.



\section{Related Work}
\label{sec:related_work}
The integration of PM into storage services, particularly key-value (KV) stores,
has received much attention in recent years.  
In this Section, we describe the landscape of persistent memory key/value stores.
We put these systems into four main categories based on their persistence 
designs: \emph{Fully Persistent}, \emph{Hybrid}, \emph{Mixed Indexing and Data}, 
and \emph{Checkpointing and Logging}. We further split these works into 2 sub-classes: 
a KV store newly designed for PM from ground up or a modification of an existing 
in-memory KV store to support PM. 
Table~\ref{tab:taxonomy_persistence} shows the taxonomy. 

\vspace{5pt}
\noindent{\bf Fully Persistent:} These systems choose to maintain all 
indexing structures and key-value data inside persistent memory. This is the simplest design. 
However, extensive writes incur the additional latency of PM. 
As a result, newly designed persistent KV stores focus on reducing the number
of writes to their indexing structure (caused by actions such as hash table
resizing, tree node splitting) with various optimization techniques
including level hashing~\cite{Zuo:2018:WHH:3291168.3291202}, sorted node
organizations~\cite{chen2011rethinking} and indirection~\cite{227812,
Chen:2015:PBT:2752939.2752947}. In addition, they can customize the
recovery process using unique data structures~\cite{227812} to reduce 
the number of flushes. Other newly designed fully persistent 
systems use B+-trees for indexing. They make optimizations to minimize writes, 
such as keeping nodes unsorted and merging tree nodes~\cite{Chi:2014:MBT:2627369.2627630}.

Our fully persistent implementations follow works that modify existing KV 
stores~\cite{MaratheHotStorage2017,NalliASPLOS2017}. 
These systems are designed for DRAM-based architecture and feature
fewer write optimizations. 
Debnath \emph{et al.}~\cite{Debnath:2015:RHT:2819001.2819002} study how different
DRAM-based hashing schemes perform when directly ported to PM with few optimizations. 
In WHISPER's fully persistent port of Memcached~\cite{NalliASPLOS2017}, they 
allocate the hashtable in persistent memory segments and surround all accesses to
persistent memory in durable transactions.
In Oracle's implementation of persistent Memcached, they started with a hybrid
design, but converge on a fully persistent design when they realized that
recovery without persisting related data structures, such as the slabs and LRU,
proved to be difficult. \cite{MaratheHotStorage2017}

\vspace{5pt}
\noindent{\bf Hybrid:}
Other persistent KV stores choose to keep their indexing structure in volatile
memory and their key-value data inside PM, which we refer to as {\em hybrid}. 
The benefit is that writes to the index, including extending and reorganizing indices, 
occur in memory at lower latencies. Some hybrid KV stores do not implement recovery 
logic, focusing on performance evaluation or using PM to increase capacity. 
WHISPER's Redis~\cite{SnalliRedis} and Intel's PMEM Redis~\cite{intelpmemRedis} 
replace volatile key-value data allocations with persistent allocations and add basic query support. 
NVMCached~\cite{wu2016nvmcached} trades data loss for performance (reduced flushes) and 
stores the checksums of KV data in a persistent data structure called a ``zone'' 
to allow verifying data integrity upon restart. Those systems that do support 
recovery need to properly recover indexes, which are volatile and not crash consistent.
Without a persisted indexing structure, Hybrid KV-stores need a way to access
their persistent data upon restart in an organized manner. Strategies include allocating 
ranges in segments~\cite{KourtisFAST2019} and using auxiliary data
structures such as persistent slabs~\cite{LenovoMemcached}. Our Hybrid implementation 
leverages the fact that the PMDK allocator keeps track of all persistent memory allocations 
and uses its exposed iterator interface to reconstruct hash table upon restart.

\begin{table*}[t]
  \footnotesize
  \centering
  \begin{tabular}{|m{1.8cm} | m{2cm} | m{3.4cm} | m{1.1cm} | m{1.1cm} | m{1.4cm}| m{1cm} | m{1.2cm}|} 
 \hline
  {\bf Work} & {\bf Type of System} & {\bf Design Category} & {\bf Regular Perf.} & {\bf Tail Latency} & {\bf Recovery Perf.} & {\bf Data Loss} & {\bf Scalability} \\
 \hline
 \cite{chen2011rethinking,Chi:2014:MBT:2627369.2627630,Debnath:2015:RHT:2819001.2819002} & Newly Designed & Fully Persistent &  \checkmark &  &  &  & \\
 \hline
 \cite{Chen:2015:PBT:2752939.2752947} & Newly Designed & Fully Persistent & \checkmark &  &  & \checkmark\text{*} & \\
 \hline
 \cite{Zuo:2018:WHH:3291168.3291202} & Newly Designed & Fully Persistent & \checkmark &  &  &  & \checkmark \\
 \hline
 \cite{227812} & Newly Designed & Fully Persistent & \checkmark & \checkmark & \checkmark &  & \checkmark\\
 \hline
 \cite{KourtisFAST2019} & Newly Designed & Hybrid & \checkmark & \checkmark &  &  & \checkmark \\
 \hline
 \cite{wu2016nvmcached} & Newly Designed & Hybrid & \checkmark &  &  & \checkmark\text{*} & \\
 \hline
 \cite{bailey2013exploring} & Newly Designed & Mixed & \checkmark &  & & \checkmark\text{*} & \checkmark\\ 
 \hline
 \cite{Kim:2017:PHP:3126908.3126943} & Newly Designed & Mixed & \checkmark &  & \checkmark  & & \\
 \hline
 \cite{Xia:2017:HHI:3154690.3154724} & Newly Designed & Mixed & \checkmark &  & \checkmark & \checkmark  & \checkmark\\
 \hline
 \cite{yang2015nv} & Newly Designed & Mixed & \checkmark &  & \checkmark & \checkmark\text{*}  & \checkmark\\
 \hline
 \cite{Oukid:2016:FHS:2882903.2915251} & Newly Designed & Mixed & \checkmark &  & \checkmark & \checkmark\text{*} & \checkmark\\
 \hline
 \cite{RedisFlash,intelpmemRedis,SnalliRedis} & Modification & Hybrid: Redis & \checkmark & & & & \\
\hline
\cite{PMEMRedis,RedisFlash} & Modification & Mixed: Redis & \checkmark &  & & & \\
 \hline
 \cite{LenovoMemcached} & Modifcation & Hybrid: Memcached & &  &  & & \\
 \hline
 \cite{MaratheHotStorage2017} & Modification & Fully Persistent: Memcached  & \checkmark & & \checkmark\text{*} & & \checkmark\\
 \hline
 \cite{NalliASPLOS2017} & Modification & Fully Persistent: Memcached & \checkmark &  &  & & \\
 \hline
 \cite{libpmemlog-AOF} & Modification & Checkpoint/Logging: Redis & \checkmark &  & \checkmark & & \checkmark \\
 \hline
\end{tabular}
  \caption{Taxonomy of existing persistent memory KV stores by design tradeoffs they have explored. 
  $^*$ No evaluation but mentions in writing.}
  \label{tab:taxonomy_tradeoff}
  \vspace{-0.1in}
\end{table*}

\vspace{5pt}
\noindent{\bf Mixed Indexing and Data:}
These systems maintain data and/or indexes in the mixture of volatile and persistent 
memory~\cite{Kim:2017:PHP:3126908.3126943,Xia:2017:HHI:3154690.3154724,Oukid:2016:FHS:2882903.2915251,
bailey2013exploring}. For example, some systems split indexes so that some
parts of the index are in volatile memory and other parts of it are in
persistent memory. This differs from the hybrid design as hybrid is purely
volatile indexing and purely persistent data. A common case keeps the leaf
nodes of a B+-tree in persistent memory and the interior nodes in volatile
memory \cite{Oukid:2016:FHS:2882903.2915251}.
PapyrusKV~\cite{Kim:2017:PHP:3126908.3126943} stores local MemTables (in-memory 
data structure that stores KV pairs) in volatile memory and SSTables (Sorted String 
Table, which stores immutable KV Pairs after MemTables have reached max capacity) 
in PM as a form of indirection to increase performance in a distributed setting. 
HiKV~\cite{Xia:2017:HHI:3154690.3154724} keeps a hash index in volatile memory for 
high-frequency updates and a B+-Tree index in persistent memory. Echo~\cite{bailey2013exploring} 
is a PM KV store that has threads store data in local hashtable stores before being 
added to a queue to be added to a global persistent store. Redis Lab's Redis on 
Flash~\cite{RedisFlash} stores its keys, dictionary, and hot values inside DRAM 
while storing its warm values on SSDs. Crash recovery in Redis-on-Flash relies on 
Redis' disk-based snapshots. PMEM-Redis~\cite{PMEMRedis} places values in persistent 
memory that are larger than an PM threshold size while keeping smaller values in memory. 
NVTree~\cite{yang2015nv} is a B+-Tree that only enforces consistency on leaf nodes (critical data) 
and does not guarantee consistency for inner nodes, but keeps nodes in PM.

\vspace{5pt}
\noindent{\bf Checkpointing and Logging:} These systems keep all data in volatile 
memory but maintain
a replication medium in persistent memory. In-memory KV stores, such as Redis,
use checkpointing and logging to have some form of persistence.  Placing the
logging file or snapshot in persistent memory improves performance, because
writes to persistent memory have much smaller latencies than writes to disk 
or SSDs. Bullet~\cite{Huang:2018:CPG:3277355.3277448} uses Cross-Referencing 
Logs that record both the key-value data and ordering dependencies among 
records to allow proper recovery. libpmemlog-AOF-Redis~\cite{libpmemlog-AOF} 
supports recovery using a persistent Append-Only File that logs every write operation 
and replays them upon restart.

\vspace{-0.05in}
\section{Motivation and Scope of Investigation}
Our taxonomy in Table~\ref{tab:taxonomy_persistence} demonstrates that there exist 
a variety of design choices for KV stores with persistent memory. It also shows that 
there is no consensus about how to design such systems. It is crucial to understand 
the trade-offs implied by different designs. Thus we further organize these related 
work based on the system properties that are evaluated as shown in Table~\ref{tab:taxonomy_tradeoff}. 
We can see that almost all systems measure regular performance (throughput and
latency) and many also evaluate scalability. However, other properties such as recovery 
performance, tail latency, data loss are much less often examined. These
properties are as important for a real-world persistent-memory KV store 
operating in production.


Moreover, while many KV stores have been built from the ground up and customized 
for persistent memory such as uDepot~\cite{KourtisFAST2019} and CCEH~\cite{227812}, 
it is increasingly common for developers to add persistent memory support to an 
existing in-memory KV store. We refer to this process as {\em porting}. Porting builds 
on the battle-tested maturity of the existing system, inherits its 
operational properties, minimizes code complexity, and eases adoption. Porting
also presents unique challenges in properly integrating the modifications with existing
code that is originally designed for DRAM. Unfortunately, as Table~\ref{tab:taxonomy_tradeoff} 
shows, porting is not well explored in current literature.

In this work, we aim to shed some light on the aforementioned gap by comparing the design 
choices of \emph{porting} an in-memory KV store to persistent memory, 
and comprehensively quantifying the trade-offs of these choices. We focused 
our efforts on fully persistent and hybrid systems because these techniques both (1) 
take advantage of the byte-addressable properties of persistent memory and (2) can be used to add 
recoverability to existing systems. Mixed systems are interesting and complex and 
typically require an entire re-design. 

A closely related work examines the difficulties in porting Memcached to
persistent memory~\cite{MaratheHotStorage2017}. They cover many
salient points in the design of a ported key/value systems, such as 
tracking persistent and non-persistent object interactions, the necessity of
using failure-atomic transactions, and the difficulty in deciding which data
structures to persist.  Their treatment is limited to Memcached and the fully 
persistent design only. Their evaluation also focuses on regular performance and 
scalability (recovery is mentioned to be ``instantaneous'' but not quantitatively 
evaluated).

\vspace{-0.05in}
\section{Design}
\vspace{-0.05in}
\label{sec:design}
We present two designs for porting Redis and Memcached to use PM
for fault-tolerance. The first, {\em hybrid} design stores all key-value data 
in PM and maintains indexing structures in volatile memory.  The second, 
{\em fully persistent} maintains all data structures in PM, including all indexing 
and bookkeeping structures. All other aspects of the two designs are made similar in 
order to isolate and highlight the effects of this fundamental 
difference. For example, both use the same PM programming primitives. 
The designs are also minimal. We inherit as much as we 
can from the original implementation in an attempt to preserve the properties of 
the original systems.

We consider and later evaluate (Section~\ref{sec:eval}) a number of properties: 
(1) operational throughput: the performance while executing \texttt{GET}/\texttt{SET}/\texttt{UPDATE}/\texttt{DELETE} 
queries; (2) recoverability: if the system can properly recover the data with 
consistency and minimal loss; (3) recovery performance: how fast the systems can
recover the persisted data and reconstruct in-memory structures from it upon restart 
to continue operations; (4) tail latency: performance influenced by data structure 
re-organizations, background tasks \emph{etc.}; (5) concurrency effect: would the 
ported systems handle concurrency properly as before and scale; (6) development effort: 
the extensiveness and difficulty of making the modifications to the system.


\vspace{-0.1in}
\subsection{Challenges and Design Trajectory}
A key challenge in both designs was determining which data structures to
persist. Making one data structure nonvolatile can create a large number of
dependencies. Other variables that are referred to either (1) may be made 
persistent as well; (2) may be kept volatile and need to be recovered on restart. 
For example, when persisting Memcached's hashtable, we also had to persist the 
internal slab pointers that Memcached uses to track allocations. We also had to 
persist global variables that contain system metadata, such as the hash power and 
hashsize. However, one needs to be careful not to persist any unnecessary data structures 
or variables, because persistent memory writes reduce performance.  Our strategy was 
to first choose a primary data structure to persist and then trace all of its 
internal variables to see if they also needed to be persisted or could be recovered. After this, 
we used testing and static analysis to determine that the state was either 
persistent or correctly recovered on restart, and then repeat the process. 

It turns out that the two choices lead by induction to our two designs.  When
you decide to persist dependent variables, you create further dependencies that 
also need to be persistent. When you decide to recover a data structure instead, 
its dependencies must also be recovered.  We do note that even in the {\em fully 
persistent} implementation, we keep as many variables volatile as possible, 
\emph{e.g.} cache state, for performance reasons. The choice of what to persist was 
much deeper and more complex than we expected.

Another challenge is to recover persistent pointers, because persistent
memory regions are mapped into different addresses on each system 
instantiation. The first option is to change all pointer references in the code to
address memory using a persistent memory offset.  The other choice is to
update/rewrite all pointers during recovery.  Systems like Memcached that
rely on a slab allocator already use compound pointers and translate more
easily to persistent memory.

\vspace{-0.1in}
\subsection{PMDK Library}
Unlike many related work that use the low-level \texttt{clwb} and \texttt{sfence} 
instructions to program persistent memory, our implementation uses Intel's 
Persistent Memory Development Kit (\textsf{PMDK}). \textsf{PMDK} provides 
a set of libraries with high-level programming constructs and APIs for developers 
to use. The libraries build on the DAX (direct access) support from the OS that 
allows applications to use accesses persistent memory device as memory mapped files.
We chose this library persistence programming model for two main reasons.
First, the PMDK APIs are simple to use and greatly ease the porting efforts
as one does not need to reason about persistence at the cache line granularity. 
Second, cache line flush and fence alone are not enough to ensure recoverability. 
In existing KV stores, handling a single request like \texttt{INSERT} typically 
involves modifying multiple data structures across a series of complex operations. 
These modifications need to be atomic.


Particularly, we leverage PMDK's \textsf{libpmemobj} library which
provides a transactional object store for persistent memory management to
ensure proper data consistency within the persistent memory mapped file. 
Developers define the transaction region and call the \textsf{libpmemobj}'s 
transactional functions (\texttt{pmemobj\_tx\_alloc}, \texttt{pmemobj\_tx\_add\_range}, 
etc.). Transactions can be nested. Writes within the transaction will be flushed at 
the end of the transaction. In the case of an unexpected crash, \textsf{libpmemobj}
uses an undo log to properly undo all persistent changes that had occurred within 
transactions.

\begin{figure}[t]
  \centering
  \includegraphics[width=3in]{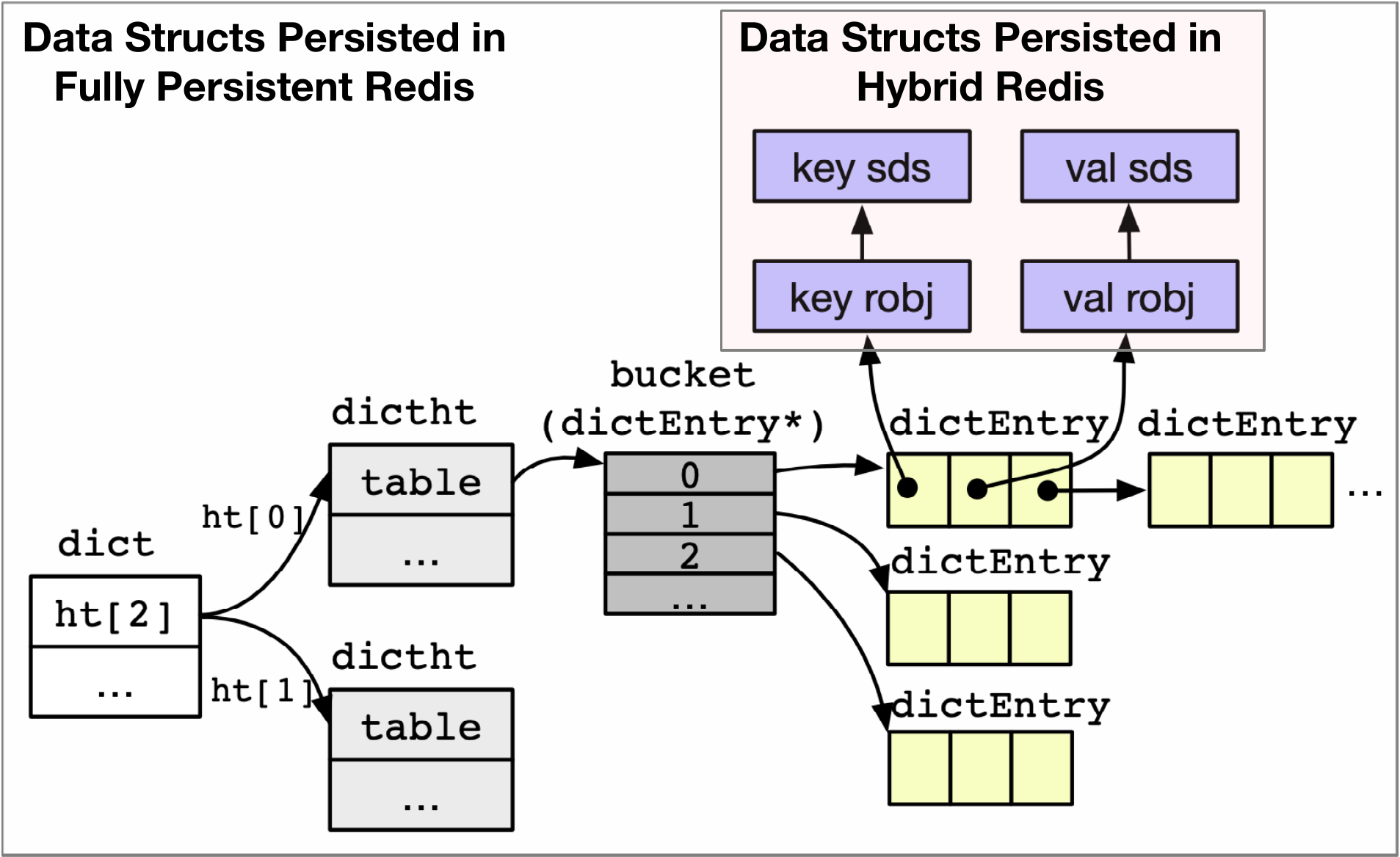}
  \caption{Fully Persistent Redis vs. Hybrid Redis, what core data structures in Redis are made persistent in each design.}
  \label{fig:full_hybrid_redis_overview}
  \vspace{-0.1in}
\end{figure}

\vspace{-0.1in}
\subsection{Fully Persistent Redis}


Fully Persistent Redis ports Redis' data and indexes in persistent memory in order to 
minimize interactions between volatile and nonvolatile data.
We replace all relevant volatile allocations with 
\textsf{libpmemobj}'s persistent alloc function.
These include the Simple Dynamic String (\texttt{sds}), Redis Object (\texttt{robj}), 
Dictionary Entry (\texttt{dictEnty}), and the hashtable (\texttt{dict}, \texttt{dictht}, 
and \texttt{bucket}). 
Figure \ref{fig:full_hybrid_redis_overview} shows the hierarchy of data structures and indicates 
that they are all placed in persistent memory.

\begin{figure}[t]
  \centering
  \includegraphics[width=3in]{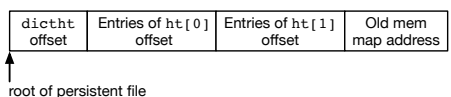}
  \vspace{-0.1in}
  \caption{Fully Persistent Redis' stored offsets at root}
  \label{fig:pers_offsets}
  \vspace{-0.15in}
\end{figure}

Although placing hashtables in persistent memory streamlines the recovery process, 
we still need to remap all persistent pointers. Recovery requires reattaching the persistent hash tables. 
Upon application start, the {\sf libpmemobj} library creates or opens a memory
mapped file, called a pool, that contains a root virtual address. A new memory 
mapped address will be assigned for the root, effectively displacing 
all the previously saved direct, persistent pointers. Thus, all Redis' hashtables 
and datastructures are virtual addresses with respect to the last system startup
and become invalid when the system is restarted. To make these pointers valid,
we keep track of the old address where the persistent memory device was mmaped,
and use the address translation with respect to the following formula: 
\( new\_pointer = old\_pointer - old\_mmap\_address + new\_mmap\_address \).
\noindent Using this formula, we walk through the hashtables and reconstruct 
all the pointers. During a restart, Fully Persistent Redis iterates over a hashtable 
entry and validates all of its pointers to key value data before reconstructing the ``next'' hashtable entry
pointer and moving on to that entry. Figure~\ref{fig:pers_offsets} shows that the
base Hashtable offset, the base entry of \texttt{HT[0]}'s offset, the base entry of
\texttt{HT[1]}'s offset and the old memory mapped address are all stored contiguously at
the root of the persistent file and are updated whenever they are modified
during Redis' operational time. 

One way of circumventing this issue is to modify all of Redis' data structures
to just use offsets rather than pointers. In this manner, you no longer have to
reconstruct pointers every time you restart as your offsets remain constant.
However, this requires significant coding effort as you have to replace every
single pointer and pointer reference with offsets and offset memory access
helper functions respectively.

As a whole, Fully Persistent Redis drastically improves Redis' recovery
performance at the cost of operational latency.
Fully Persistent Redis slows hashtable modifications and accesses owing to persistent
memory's larger write latency. The main drop in performance occurs when the 
system resizes the hashtable as the key-value items grow.
Another problem that we encountered with Fully Persistent Redis was the fact that 
we had to make the random \texttt{hashseed} for Redis' hash function constant 
across multiple runs of Redis so that the hash function is stable across restarts.


\vspace{-0.1in}
\subsection{Hybrid Redis}
Hybrid Redis maintains the indexing hashtables (\texttt{dict}, \texttt{dictht}) 
in volatile memory and only stores key-value data (\texttt{robj} and \texttt{sds}) 
inside persistent memory (Figure \ref{fig:full_hybrid_redis_overview}). By making 
the bare minimum amount of data and metatadata persistent, Hybrid Redis greatly 
reduces the number of writes and allocations to persistent memory and thus 
improves the operational performance. This improvement comes at the expense of 
a longer and more complex recovery.

For recovery, Hybrid Redis iterates over persistent key and value data and rebuilds
the volatile hashtables. The challenge is to determine how to restore all the data
across restarts without the aid of a persistent hashtable. On restart, the 
\texttt{robj} and \texttt{sds} data are available in persistent memory at new 
addresses.  But the hashtable and \texttt{dictEntry}'s are empty.
We initially tried to use a new auxiliary data structure called \texttt{bookKeeper} 
to track allocations of key-value data.  However, this approach subverts the design, 
because it makes additional writes to persistent memory on every allocation.
These maintenance costs cancels out the benefit of the hybrid design.

\begin{figure}[t]
  \centering
  \includegraphics[width=2.7in]{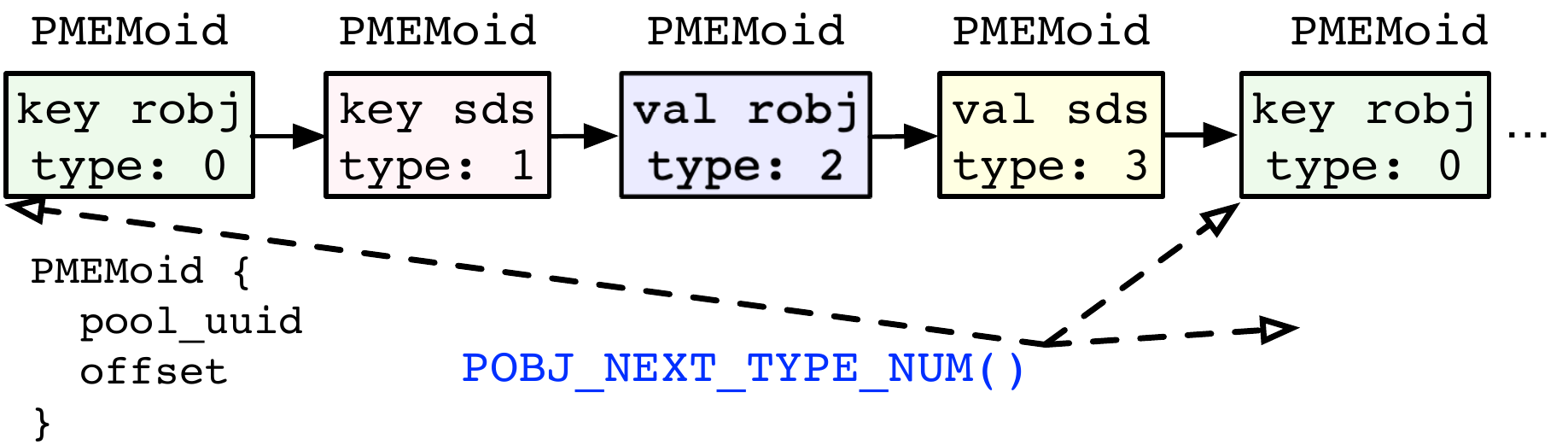}
  \caption{\textsf{libpmemobj} allocator' iterator interface allows enumeration of persistent memory allocations by assigned type numbers.}
  \label{fig:pmdk_type_iteartor}
  \vspace{-0.1in}
\end{figure}

We then determined that {\sf PMDK} tracks allocated objects and we can use this record
to iterate over allocations to discover key value data on restart.
The {\sf libpmemobj} maintains a linked list structure that tracks 
persistently allocated data across restarts. {\sf libpmemobj}'s
allocator also allows one to tag an allocation with a type enumeration (typenum). 
Redis allocates individual keys and values.  Hybrid Redis makes these allocations in
persistent memory and tags them with the object type (\texttt{key robj}, 
\texttt{key sds}, \texttt{val robj}, \texttt{val sds}) (Figure \ref{fig:pmdk_type_iteartor}).


\begin{figure}[t]
\centering
\includegraphics[width=2in]{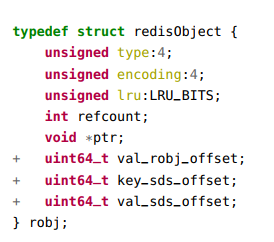}
\vspace{-0.15in}
\caption{Modified Redis \texttt{robj} to store persistent memory addresses.}
\label{fig:robj_modification}
\vspace{-0.1in}
\end{figure}

We augment the Redis \texttt{robj} data structure to contain recovery information 
for a key-value pair. Because there are no guarantees on a fixed order for allocations, 
we modified the \texttt{robj} structure to contain all of the addresses for key value 
data that is necessary for reconstructing the key value pair upon restart. 
Figure \ref{fig:robj_modification} shows the added fields that record the persistent memory offsets 
from the base.  On restart, we use the allocator link list to identify key objects and then 
use the offsets in each key to locate the related key data and value object and data.  
The discovered object is inserted into the hashtable.  After traversing all allocated objects,
the hashtable is reconstructed so that it indexes same contents as it did before shutdown or failure.

\vspace{-0.1in}
\subsection{Fully Persistent Memcached}
\begin{figure}[t]
  \centering
  \includegraphics[width=3in]{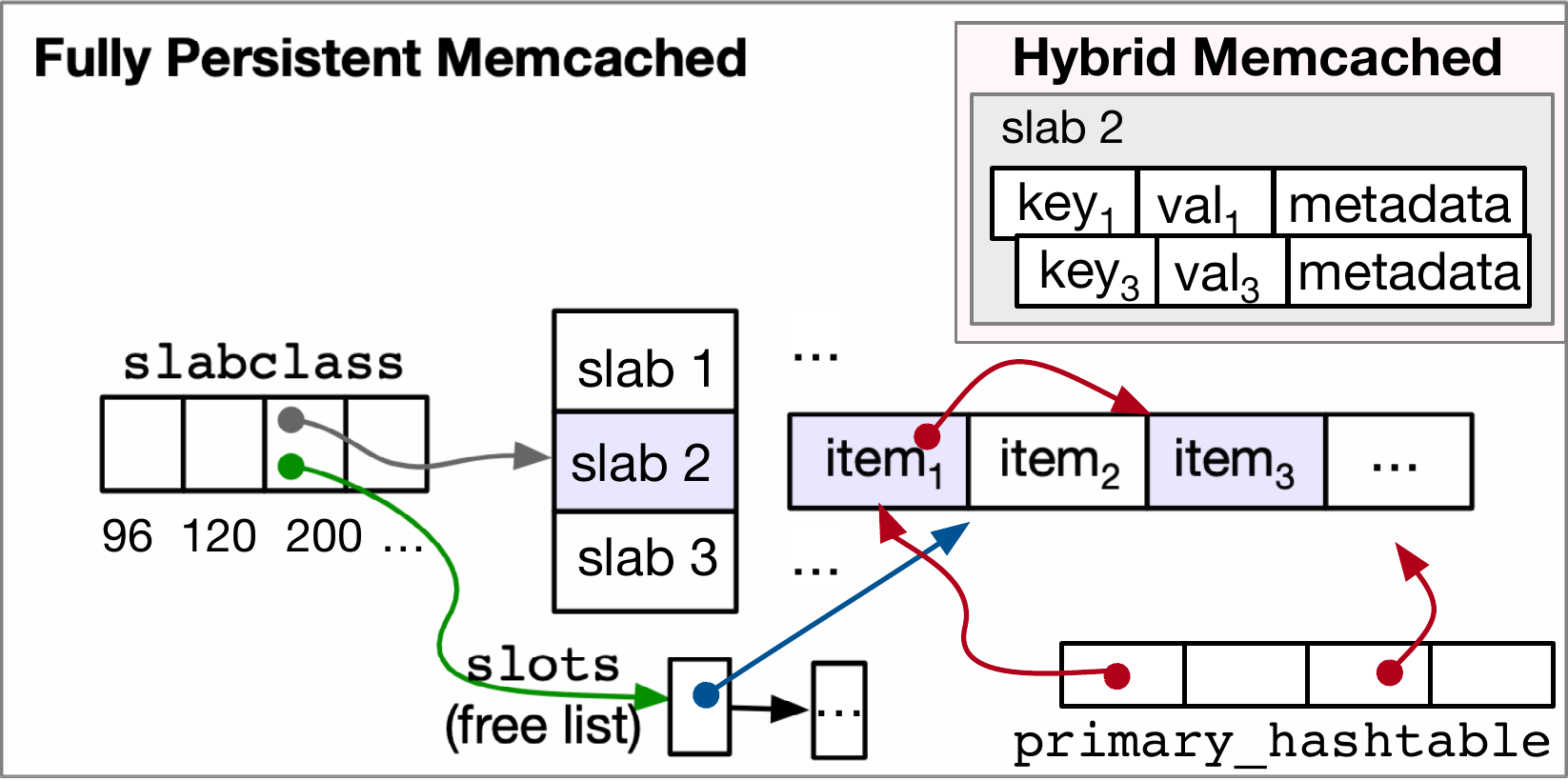}
  \caption{Fully Persistent Memcached vs. Hybrid Memcached}
  \label{fig:pers_memached}
  \vspace{-0.1in}
\end{figure}

For the most part, Fully Persistent Memcached follows the same design principles as 
Fully Persistent Redis. We maintain the hashtable, all linked lists, and key-value 
data structures inside persistent memory (Figure \ref{fig:pers_memached}) in order 
to ease recovery at the expense of reduced operational performance. 

The major difference is that memcached uses a slab allocator so that we allocate and 
manage key-value data on a slab by slab basis rather than at a fine granularity.
Memcached allocates chunks of varying size.
Variable length data is placed contiguously in the appropriate slab until Memcached
has to allocate a new slab. To access the key and value strings inside
of slabs, Memcached uses a set of mathematical macros that calculate the
address of the strings using the offsets from the given item pointer.
We make the allocated item slabs persistent in Fully Persistent
Memcached. This differs from Redis where we persist individual key and value objects.
This improves the performance of Memcached when compared with its persistent Redis variants,
because it amortizes the cost of allocation over multiple key-value pairs. 

The recovery process in Memcached follow that of Redis, but is slower in practice
because Memcached uses more pointers for the same number of key/value pairs.
Similar to Redis, we validate every pointer on restart. However, 
in addition to the hashtable and key-value object pointers, 
Memcached has a \texttt{slabclass} linked list and caching data structures.
We split the recovery process into two portions. First, we
walk the \texttt{slabclass} array of slab pointers,
validating each item in every slab individually. Then, we walk through the
hashtable pointers and validating the pointers for each item entry.

\vspace{-0.1in}
\subsection{Hybrid Memcached}
\begin{figure}[t]
\centering
\includegraphics[width=3in]{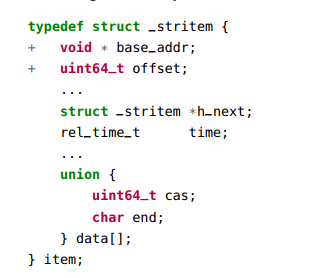}
\vspace{-0.15in}
\caption{Hybrid Memcached \texttt{item} structure modifications.}
\label{fig:memcached_mod}
\vspace{-0.1in}
\end{figure}

Hybrid Memcached maintains indexes and cache data structures in volatile memory and stores key-value data in 
persistently allocated slabs, following the slab allocation schema of Memcached.
In Hybrid Memcached, the old slab data structures are still volatile (Figure \ref{fig:pers_memached}). 
However, the key and value data are now stored in the persistent slab.
This persistent slab contains the minimum amount of information
to reconstruct its key/value pairs upon restart, including key data, value data, and corresponding metadata. 
Recovery reads all the slabs to rebuild \texttt{slabclass} and then recovers 
all keys and values in the slabs to repopulate the hashtable and cache. 
To accomplish this, we modify the \texttt{item} class and addressing macros to refer to keys and values
as a base address and offset (Figure \ref{fig:memcached_mod}).

\vspace{-0.1in}
\subsection{Dealing with Concurrency}
Since Redis is single-threaded, in our porting of Redis, we did not specially
handle concurrency. But in porting Memcached, we have to be consider 
its multithreading designs. If we access shared objects in a persistent transaction, 
acquiring a lock may be necessary (as \textsf{PMDK} transaction itself does not provide 
isolation). Fortunately, the original Memcached has properly synchronized its 
slab allocations and item modifications. So we need not add much additional 
synchronization code. We used locks mainly when we are modifying the global offsets 
that we added to save in the persistent items. Without these locks, we would experience 
race conditions that affect recoverability. We also prevent the added locks from affecting 
scalability by keeping the critical sections small (computing the new offsets
and saving to a local variable) and performing persistent I/O outside the 
critical section.

\section{Evaluation}
\label{sec:eval}
We evaluate the systems with the goals of comparing the hybrid and fully-persistent 
designs for both Redis and Memcached and examining six main measures 
(Section~\ref{sec:design}): operational throughput and latency, tail latency, 
recoverability, recovery performance, concurrency effect, and development effort.
We also compare our implementations with the original implementation and 
several open-source porting efforts.

We run a custom benchmark that isolates the overhead of persistent memory and hashtable 
reorganization during bulk insertions. We also run YCSB~\cite{Cooper2010} benchmarks to characterize 
system performance under various workloads: A (50/50 Reads and Writes), 
B (95/5 Reads and Writes), C (Read Only), D (Read Latest), and F (Read, Modify,Write).
The keys in the custom benchmark are 4--11 bytes and the values are 5--13 bytes. We 
use the default 1KB record size in YCSB Workloads. The experiments are performed on 
a server with one 8-core Intel(R) Xeon(R) Silver 4215 CPU (2.50GHz, 11MB L3 cache), 94 GB DDR4 DRAM, and
two 128 GB Intel Optane DC Persistent Memory DIMMs. 

\begin{table}[t]
  \footnotesize
  \centering
  \begin{tabular}{@{}llll@{}}
    \toprule
      & Sequential Read & Random Read & Write \\
    \midrule
      DRAM & \SI{81.4}{\ns} & \SI{83.2}{\ns} & \SI{157.7}{\ns}\\ 
      PMEM & \SI{179.0}{\ns} & \SI{317.6}{\ns} & \SI{160.4}{\ns}\\
    \bottomrule
  \end{tabular}
  \caption{Measure latencies of the DRAM and Intel Optane DC persistent memory in our server.}
  \label{tab:hardware_latency}
\end{table}


\vspace{0.05in}
\noindent{\textbf{Hardware Performance Characteristics}:}
We start with a simple load/store latency comparison of our persistent memory 
hardware versus DRAM. We configure the Optane DC to operate in the
App Direct mode~\cite{app_direct_mode}. This mode exposes the device as a separate storage module on the
memory bus, and applications must be modified to fully take advantage of 
the device for persistence. To measure the performance characteristic of
the persistent memory versus the DRAM, we use the Intel Memory Latency 
Checker tool~\cite{intel_mlc} (\texttt{mlc}). Table~\ref{tab:hardware_latency}
shows the measured read and write latencies. The Optane DC is about 2$\times$ 
slower than DRAM for sequential loads and about 4$\times$ slower for random loads. 
Store latency for Optane DC is only slightly slower than DRAM. These results are 
on par with the measurement by Izraelevitz \emph{et al.}~\cite{izraelevitz2019basic}.


\vspace{-0.1in}
\subsection{Insertion Benchmark}
\vspace{-0.05in}
\begin{figure}[t]
  \centering
  \small
  \begin{subfigure}[t]{0.22\textwidth}
    \includegraphics[width=\textwidth]{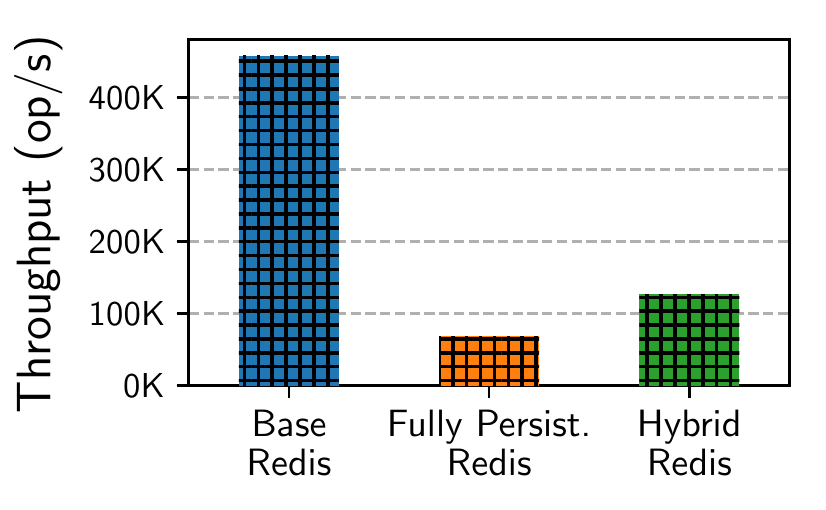}
    \caption{\footnotesize Redis Throughputs}
    \label{fig:overall_perf_redis_throughput}
  \end{subfigure}
  ~
  \begin{subfigure}[t]{0.22\textwidth}
    \includegraphics[width=\textwidth]{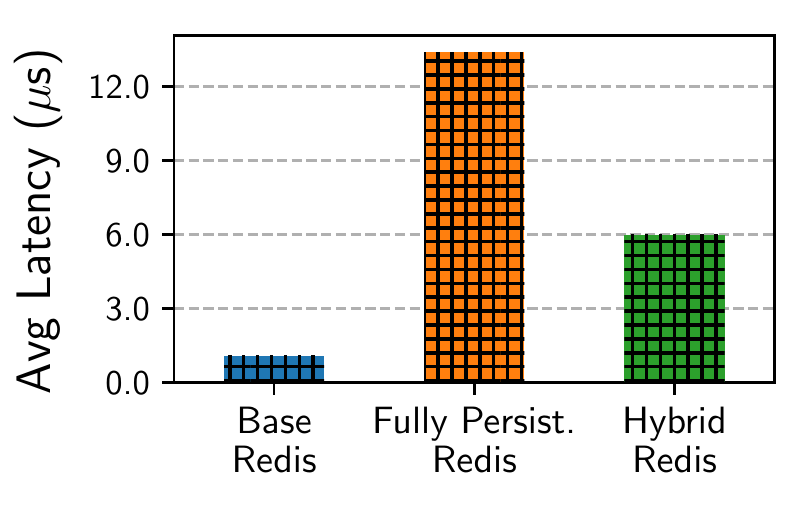}
    \caption{\footnotesize Redis Latencies}
    \label{fig:overall_perf_redis_latency}
  \end{subfigure}
  ~
  \begin{subfigure}[t]{0.22\textwidth}
    \includegraphics[width=\textwidth]{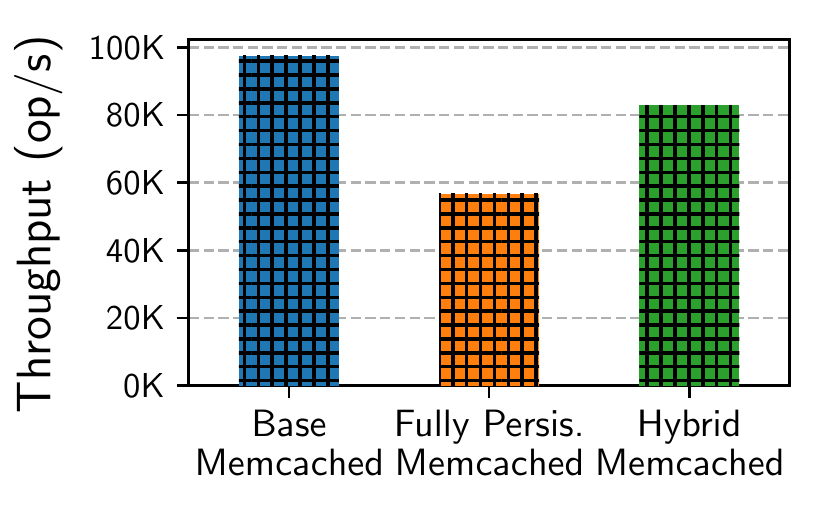}
    \caption{\footnotesize Memcached Throughputs}
    \label{fig:overall_perf_memcached_throughput}
  \end{subfigure}
  ~
  \begin{subfigure}[t]{0.22\textwidth}
    \includegraphics[width=\textwidth]{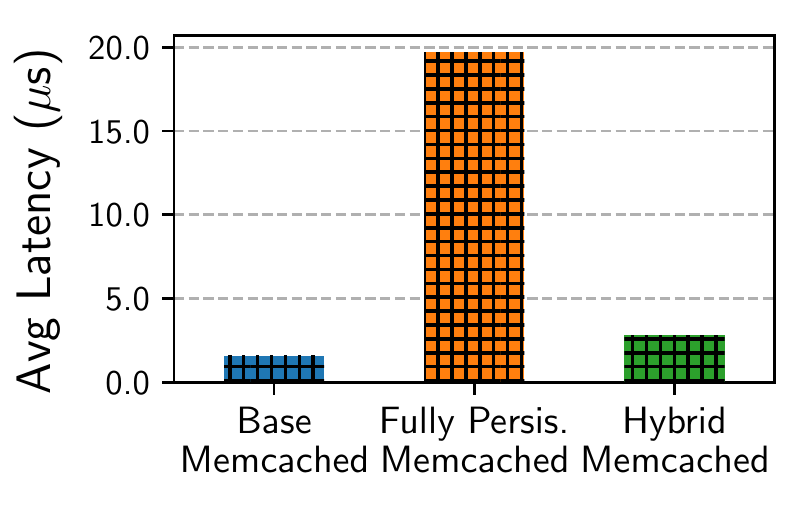}
    \caption{\footnotesize Memcached Latencies}
    \label{fig:overall_perf_memcached_latency}
  \end{subfigure}
  \vspace{-0.1in}
  \caption{Overall performance of Redis and Memcached designs.}
  \label{fig:overall_performance}
  \vspace{-0.1in}
\end{figure}

In our first experiments, we use a custom benchmark that continuously inserts 
50M unique key-value pairs into the base Redis and Memcached and the two variants 
of persistent memory Redis and Memcached. This workload reveals 
the performance structure of both writes and growing the hashtable. It best 
characterizes operational throughput and tail latency, because the slowest operations 
occur when the system reorganizes its indexing structures. 

\vspace{0.05in}
\noindent{\textbf{Operational Throughput and Latency: } Figure~\ref{fig:overall_perf_redis_throughput} 
shows the aggregate throughput for the Redis group. The persistent versions reduce 
throughput from 450,000 operations per second to below 150,000, incurring
a degradation of 3.6$\times$. This slowdown is because the persistent variants of 
Redis must write multiple offsets within a transaction, which incurs logging overhead 
and the cost of flushing writes to persistent memory when transactions commit. Hybrid 
Redis is 1.8$\times$ faster than Fully Persistent Redis because it updates the hashtable in 
DRAM rather than persistent memory. In terms of average latency, Hybrid Redis
is 2.2$\times$ better than Fully Persistent Redis as shown in Figure~\ref{fig:overall_perf_redis_latency}.

Results for Memcached in Figure~\ref{fig:overall_perf_memcached_throughput} and 
Figure~\ref{fig:overall_perf_memcached_latency} follow the same pattern as Redis: 
Hybrid Memcached is 1.45$\times$ better than Fully Persistent Memcached in 
overall throughput and 7$\times$ better in average latency. But compared to Redis, 
Memcached incurs much less overhead for persistent memory. Hybrid Memcached is only 
18\% slower that the base implementation in DRAM.  We attribute the significant
reduction in performance loss to Memcached's slab allocation amortizing allocation 
costs across multiple keys, which significantly reduces the persistent object allocations.

\begin{figure}[t]
  \centering
  \includegraphics[width=2.7in]{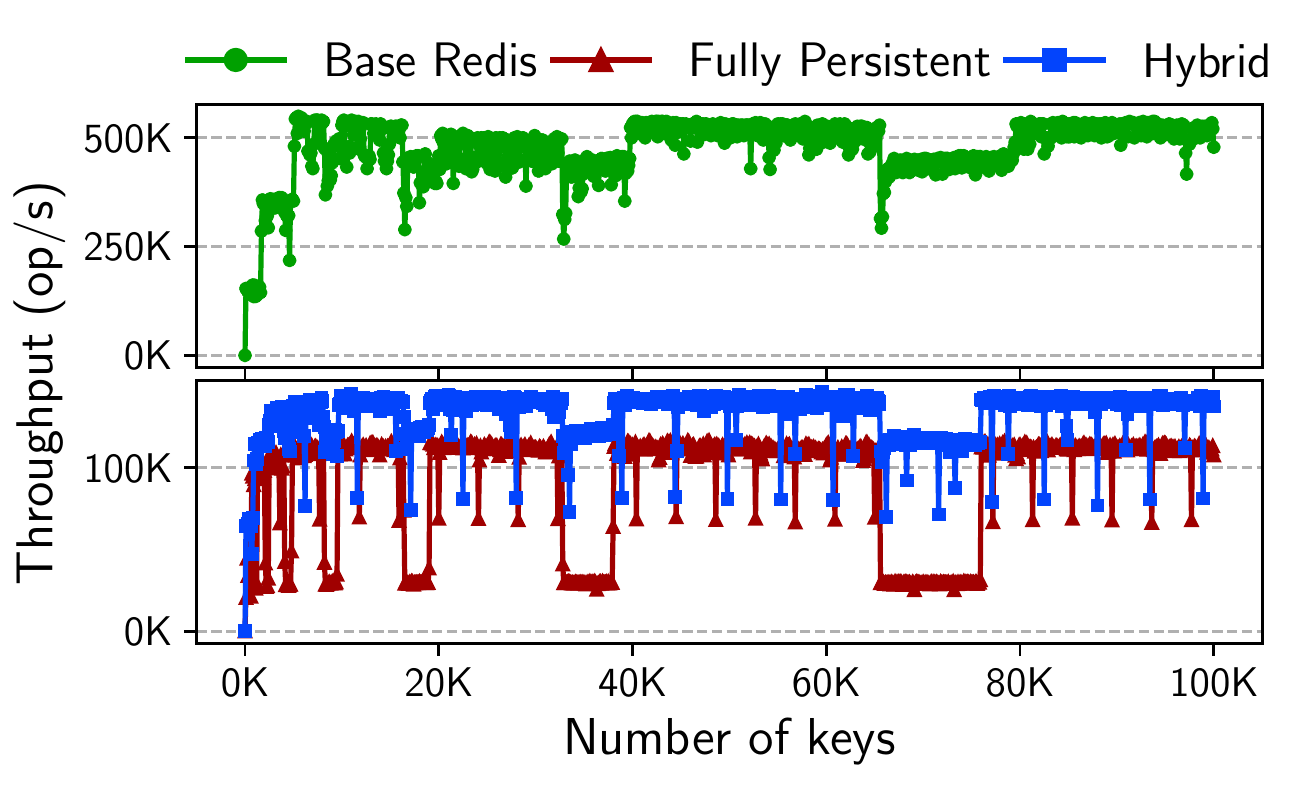}
  \vspace{-0.1in}
  \caption{Redis throughputs during the insertion experiment.}
  \label{fig:redis_throughputs_timeseries}
  \vspace{-0.1in}
\end{figure}

\vspace{0.05in}
\noindent{\textbf{Reorganization Overhead: } 
We further analyze the throughput results and find that hash table reorganization 
contributes to Hybrid designs' performance advantage.
Figure~\ref{fig:redis_throughputs_timeseries} shows the time-series throughput results for only the 
first 100,000 writes from the same insertion experiment (the remaining writes 
have similar trends). We can see that significant drops in throughputs at regular
intervals (at a power of two writes, 16K, 32K, etc.). These drops are due to hash 
table re-organizations. Interestingly, Fully Persistent Redis incurs more overhead 
(73\%) during reorganziation than Hybrid Redis (15\%). 
In Hybrid Redis the drop is lower because the writes to persistent memory of the main 
workload dominate the writes to reorganize the hashtable in DRAM.  

\begin{figure}[t]
  \centering
  \includegraphics[width=2.7in]{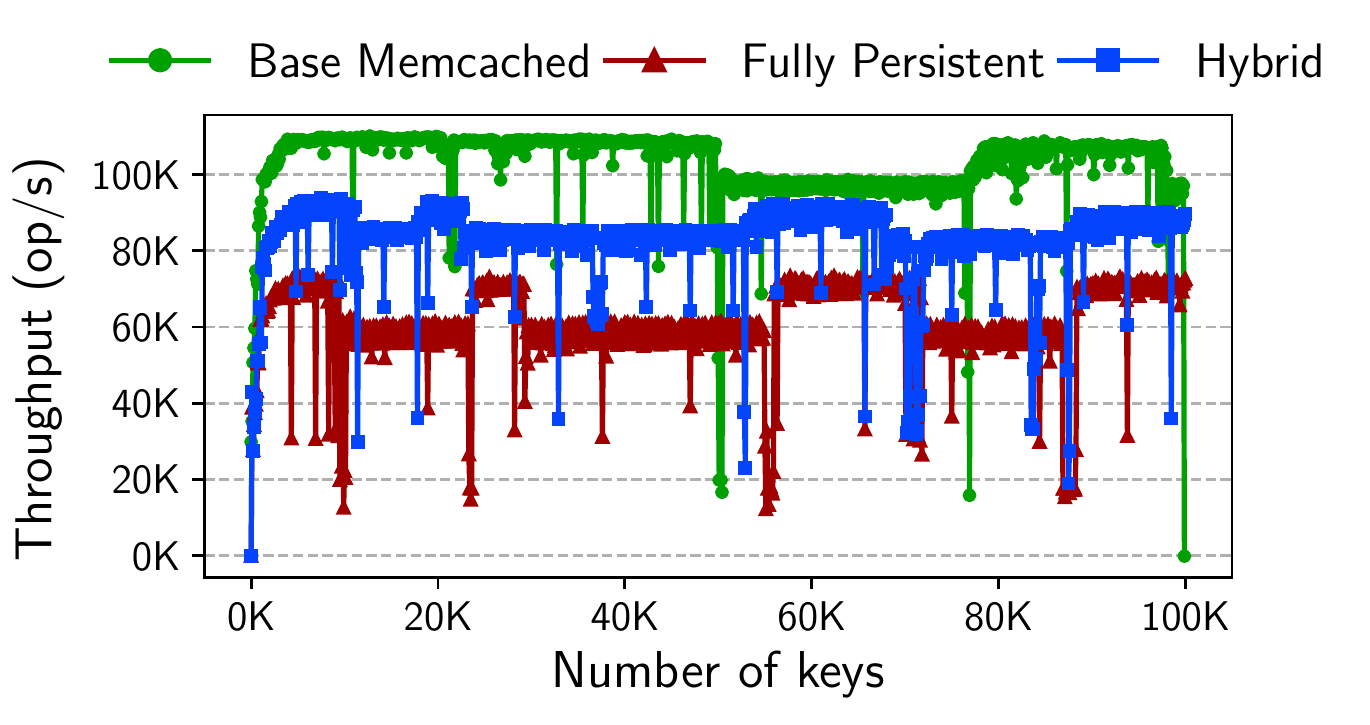}
  \vspace{-0.1in}
  \caption{Memcached throughputs during the insertion experiment.}
  \label{fig:memcached_throughputs_timeseries}
  \vspace{-0.1in}
\end{figure}

\begin{figure*}[t]
  \centering
  \begin{minipage}[t]{0.325\textwidth}
    \includegraphics[width=\textwidth]{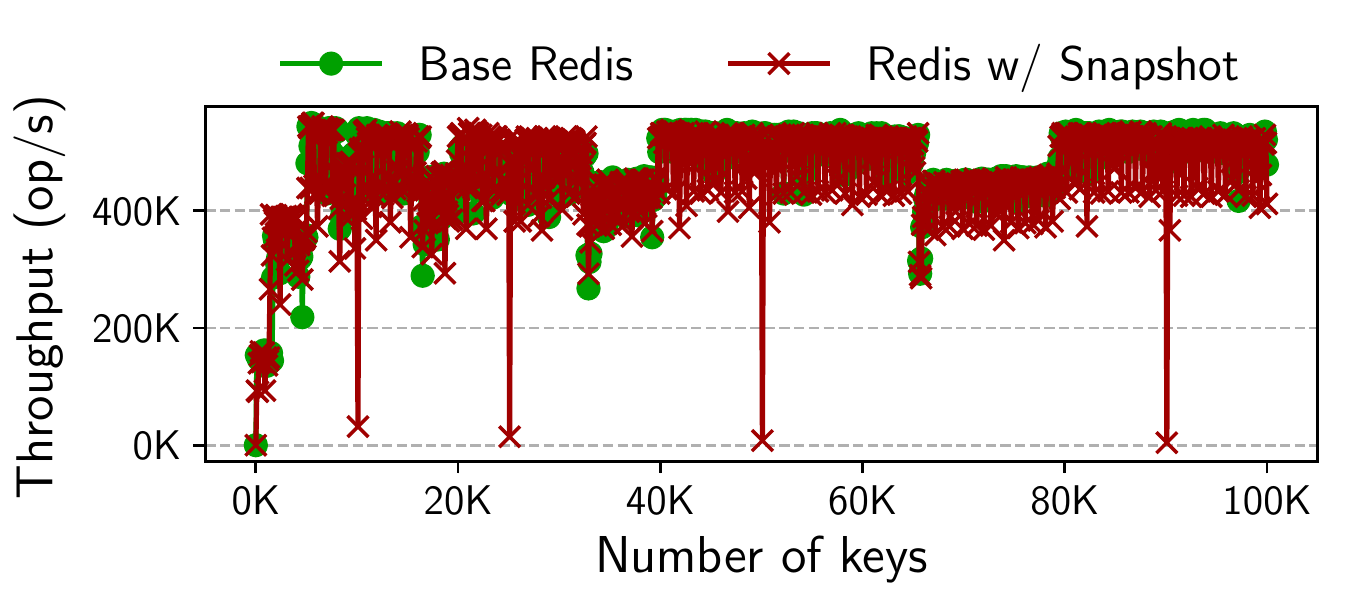}
    \vspace{-0.3in}
    \caption{Throughputs of Redis w/ snapshots.}
    \label{fig:rdb_snapshot_throughput}
  \end{minipage}
  \hfill
  \begin{minipage}[t]{0.325\textwidth}
    \includegraphics[width=\textwidth]{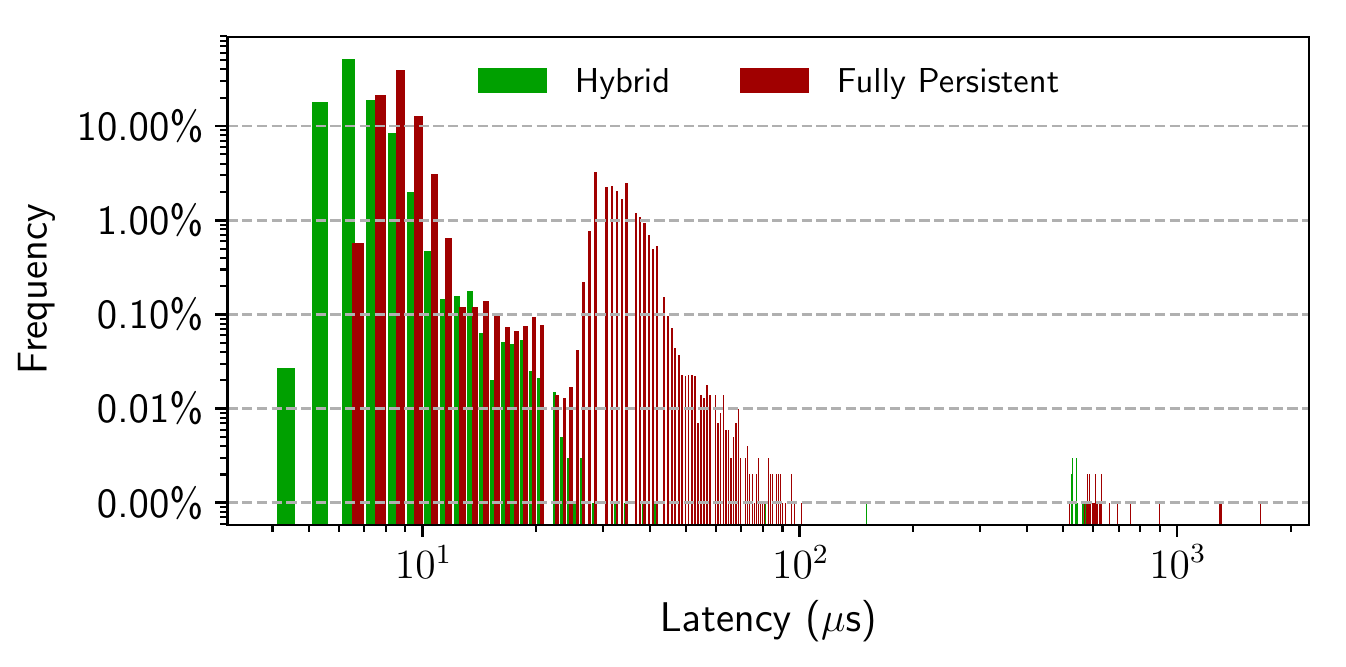}
    \vspace{-0.3in}
    \caption{Redis latency histogram}
    \label{fig:redis_latency_histogram}
  \end{minipage}
  \hfill
  \begin{minipage}[t]{0.325\textwidth}
    \includegraphics[width=\textwidth]{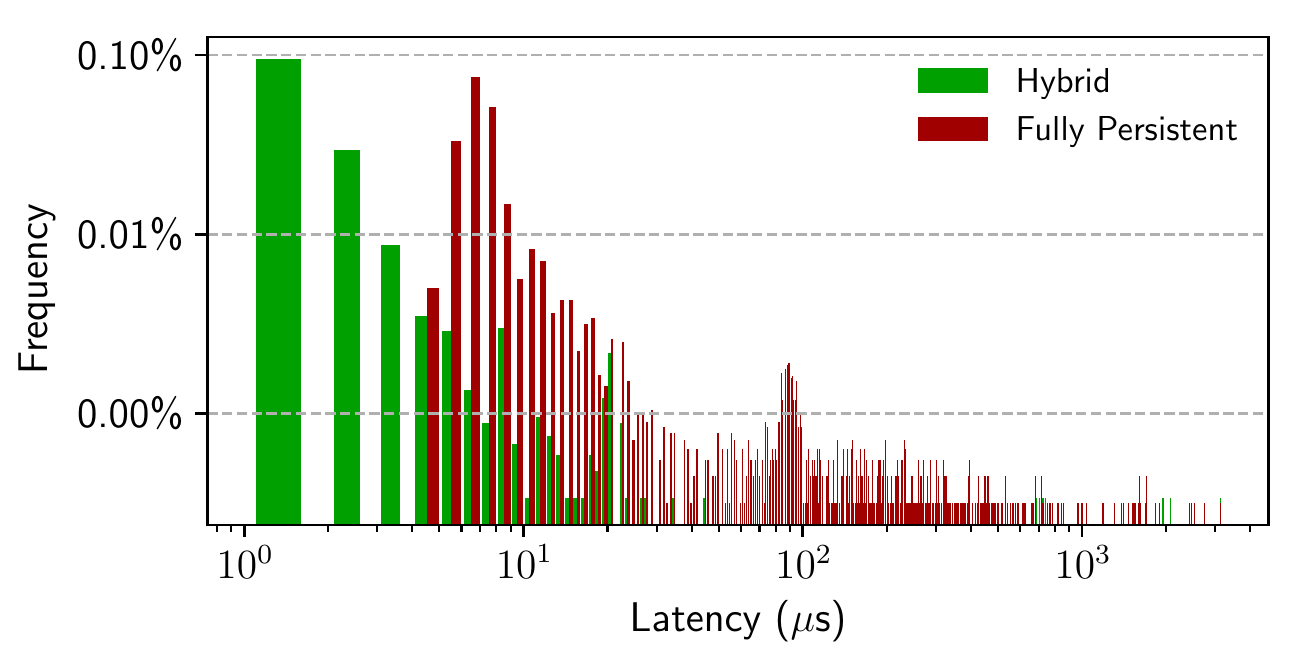}
    \vspace{-0.3in}
    \caption{Memcached latency histogram.}
    \label{fig:memcached_latency_histogram}
  \end{minipage}
  \vspace{-0.1in}
\end{figure*}

For Memcached, as Figure~\ref{fig:memcached_throughputs_timeseries} shows, 
the trend is similar to Redis: the drops in throughput due to hash table 
reorganization are 7\% for Hybrid Memcached and 19\% for Fully Persistent Memcached. 
However, Memechaed's overall drops and drop differences are smaller than Redis'. 
Memcached allocates new hashtables in slabs, which accounts for the 
better reorganization performance.

We conclude that a Hybrid design has both higher performance, about twice the 
throughput for Redis, and much more stable performance than a Fully Persistent design. 
Hashtable reorganization in persisent memory leads to substantial throughput drops, 
which we will further describe in our analysis of tail latency.

%

\begin{figure*}[t]
  \centering
  \begin{minipage}[t]{0.325\textwidth}
    \includegraphics[width=\textwidth]{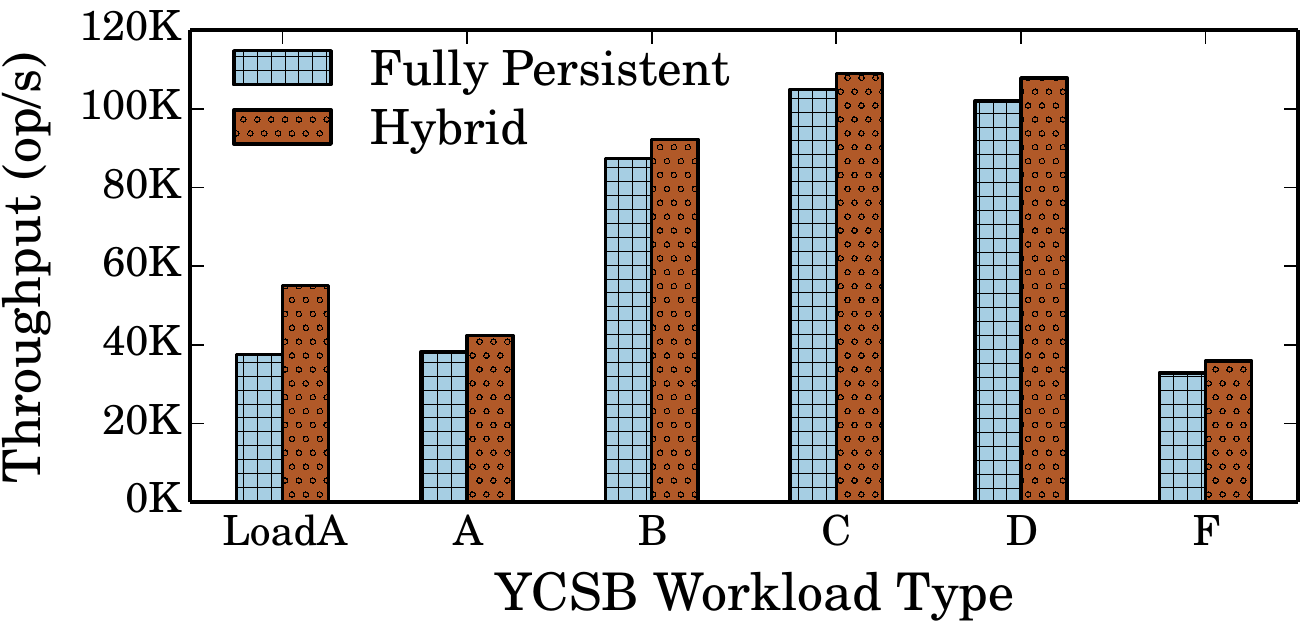}
    \vspace{-0.3in}
    \caption{Redis throughput under YCSB workload.}
    \label{fig:redis_ycsb}
  \end{minipage}
  \hfill
  \begin{minipage}[t]{0.325\textwidth}
    \includegraphics[width=\textwidth]{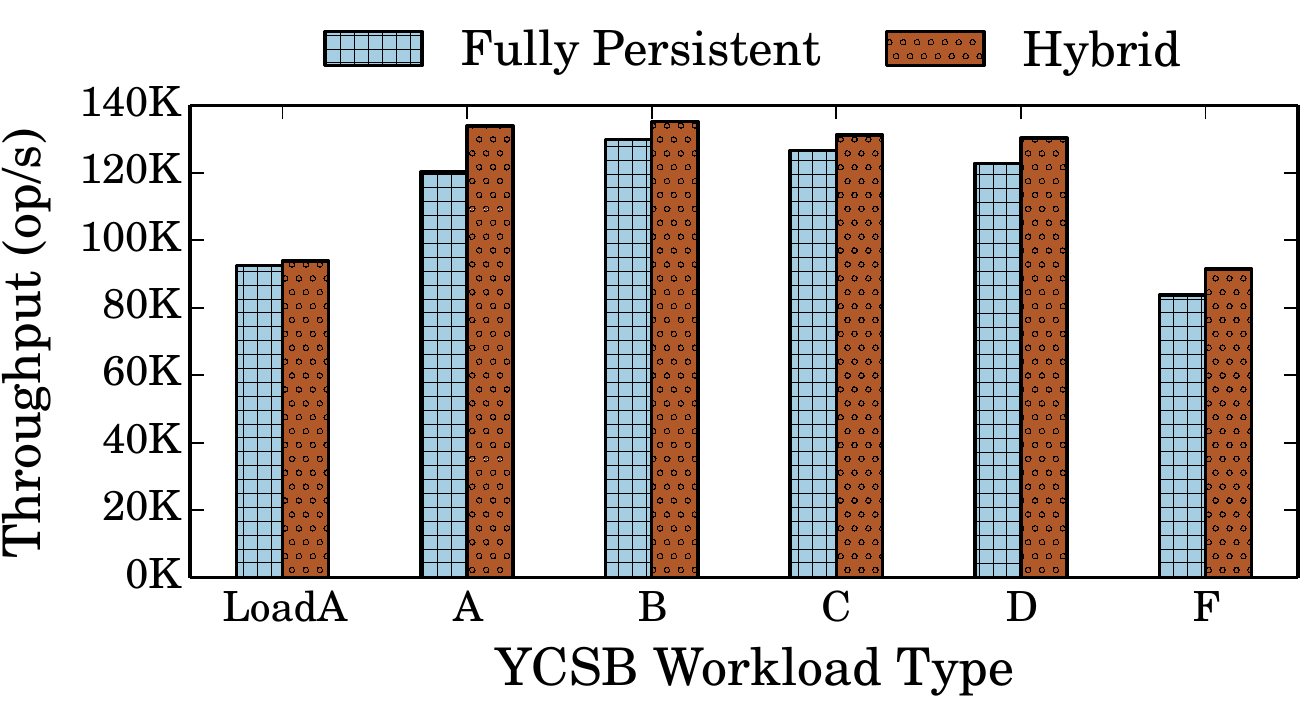}
    \vspace{-0.3in}
    \caption{Memcached throughput under YCSB workload.}
    \label{fig:memcached_ycsb}
  \end{minipage}
  \hfill
  \begin{minipage}[t]{0.325\textwidth}
    \includegraphics[width=\textwidth]{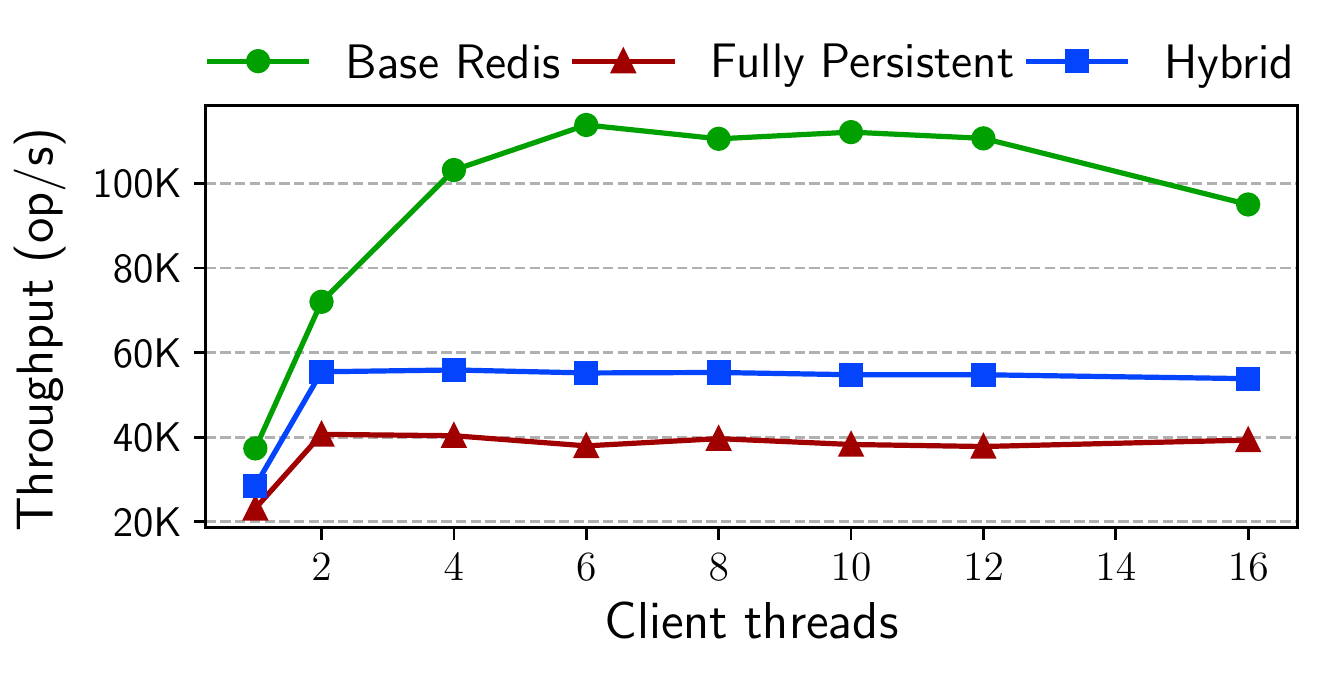}
    \vspace{-0.3in}
    \caption{Redis throughput for YCSB workloads with different number of clients.}
    \label{fig:redis_ycsb_client_throughput}
  \end{minipage}
  \vspace{-0.1in}
\end{figure*}

\vspace{0.05in}
\noindent{\textbf{Snapshot Overhead:}}
We quickly address the standard alternative for persistent storage in Redis,
which we will use as a point of comparison for throughput and tail latency. 
When users enable the RDB feature of Redis, Redis takes periodical snapshots 
to non-volatile storage.  This does not protect against data loss in the 
event of failure, but it often used by applications with weak consistency 
requirements. Figure~\ref{fig:rdb_snapshot_throughput} shows the same workload in Redis with 
and without snapshots. Both systems run at close to the same throughput except 
for the singular operations that occur at snapshot boundaries. 
%
%
%


\vspace{-0.1in}
\subsection{Tail Latency}
\vspace{-0.05in}
\begin{table}[t]
  \footnotesize
  \centering
  \begin{tabular}{@{}lllll@{}}
    \toprule
      Percentile & Base & Base+RDB & Fully Persistent & Hybrid \\
    \midrule
      50 & \SI{1}{\us} & \SI{1}{\us} & \SI{8}{\us} & \SI{6}{\us} \\ 
      90 & \SI{1}{\us} & \SI{1}{\us} & \SI{32}{\us} & \SI{8}{\us} \\
      99 & \SI{2}{\us} & \SI{1}{\us} & \SI{41}{\us} & \SI{10}{\us} \\
      99.9 & \SI{3}{\us} & \SI{3}{\us} & \SI{64}{\us} & \SI{18}{\us} \\
      99.99 & \SI{16}{\us} & \SI{14}{\us} & \SI{624}{\us} & \SI{528}{\us} \\
    \bottomrule
  \end{tabular}
  \caption{Percentile Latencies of Persistent Redis.}
  \label{tab:redis_tail_latency}
  \vspace{-0.1in}
\end{table}


%

Tail latency is important for applications to meet Service Level Agreements (SLAs)~\cite{Dean2013}.
We measure tail latency in the same insertion benchmark experiment with 50M keys. 
Table~\ref{tab:redis_tail_latency} show the latency percentiles for Redis. Both 
persistent designs have significantly worse (8$\times$ at 90th percentile) tail 
latencies compared to the base design. Hybrid Redis' tail latency at 90th percentile 
is 4$\times$ better than Fully Persistent Redis, which is attributable to its better 
performance under re-organizations. At the 99.99th percentile, the tail latencies 
have massive increase. We believe these are due to a few persistent memory operations 
(transactions) being slow. For the Redis RDB design, we see a dramatic throughput 
drop in Figure~\ref{fig:rdb_snapshot_throughput} during snapshot operations. But we 
do not see the effect on tail latencies at the 99.99\% level, because the small
number of snapshot operations during 50M insertions. Only at the 99.999\% level 
do we see the slowdown. 
The tail latencies of Memcached designs follow the same trend as Redis: the Hybrid
Memcached's tail latency at 90\% is 7$\times$ better than Fully Persistent
Memcached.

Figures~\ref{fig:redis_latency_histogram} and \ref{fig:memcached_latency_histogram} 
show the full latency distributions. In Redis, the Fully Persistent distribution 
is bi-modal with a second peak occurring half a magnitude higher than the Hybrid 
distribution's peak. In Memcached, the Fully Persistent distribution's peak
occurs a magnitude higher than the Hybird distribution's peak. The Redis histogram 
verifies that a few outlier operations take an order of magnitude more time.

\vspace{-0.1in}
\subsection{YCSB Workloads}
\vspace{-0.05in}
The YCSB benchmarks verify the performance gap between volatile and persistent memory 
and the operational throughput differences between Fully-persistent and 
Hybrid variants, showing that results apply to a variety of mixed workloads.
Figures \ref{fig:redis_ycsb} and \ref{fig:memcached_ycsb} show the results of the benchmarks
for workloads A-D and F; workload E relies on range functionality and does not apply.
These benchmarks were run with 8 clients to achieve stable and high performance.  
The performance differences between Fully Persistent and Hybrid varies between 10 and 40\%,
which is less than in the insertion benchmarks. These workloads include a mix of reads 
and writes where reads do not have transaction or allocation overheads. These workloads 
also do not grow the databases, so that they do not trigger resizing of the hashtables.

\begin{figure*}[t]
  \centering
  \begin{minipage}[t]{0.325\textwidth}
    \includegraphics[width=\textwidth]{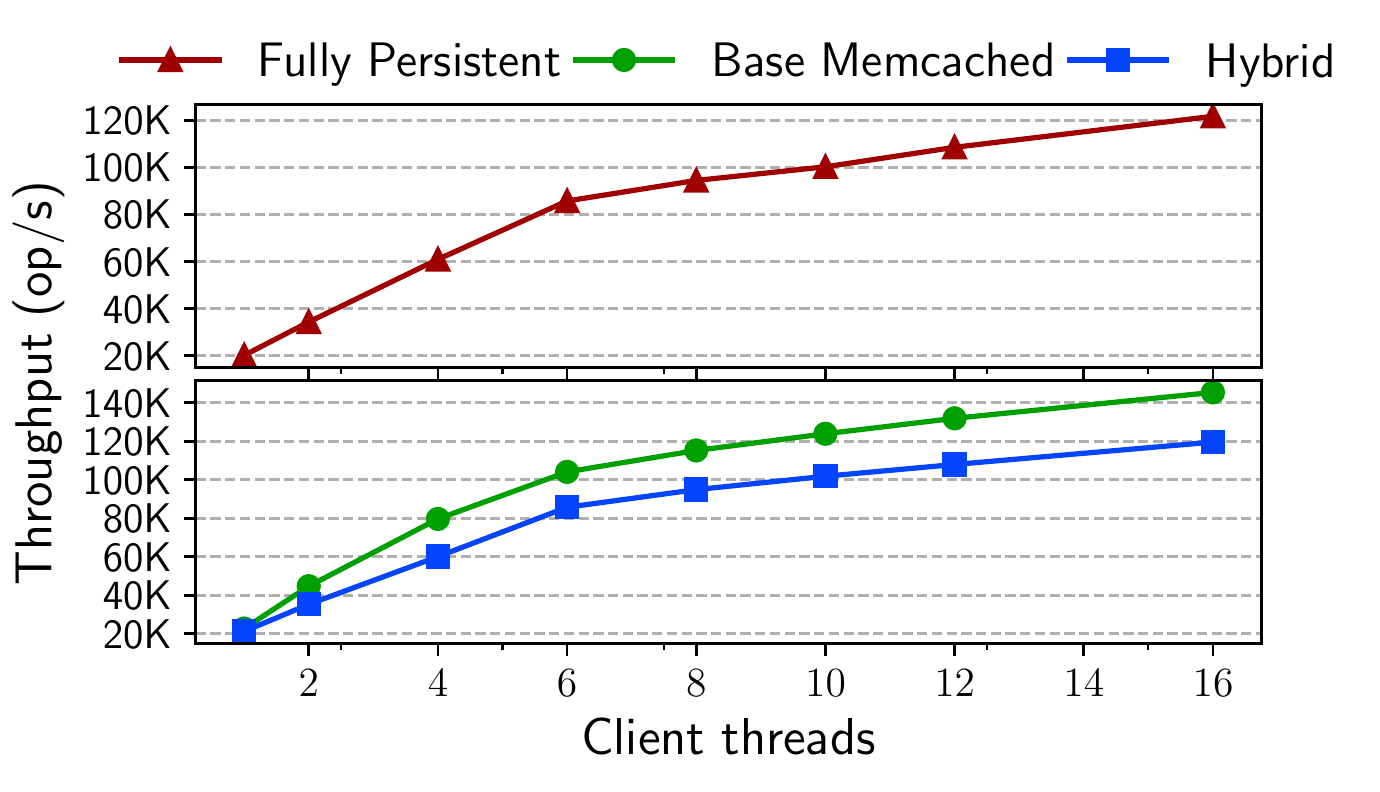}
    \vspace{-0.3in}
    \caption{Memcached throughput for YCSB workloads with different number of clients.}
    \label{fig:memcached_ycsb_client_throughput}
  \end{minipage}
  \hfill
  \begin{minipage}[t]{0.325\textwidth}
    \includegraphics[width=\textwidth]{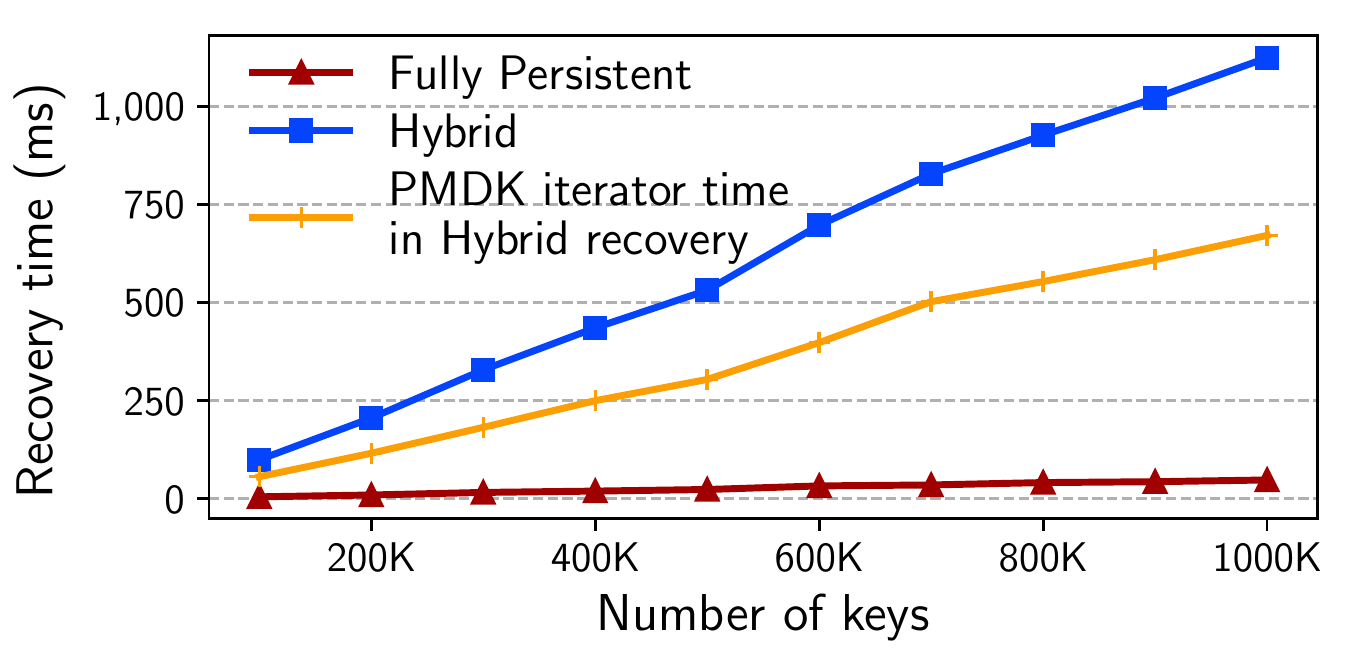}
    \vspace{-0.3in}
    \caption{Redis recovery time.}
    \label{fig:redis_recovery_time}
  \end{minipage}
  \hfill
  \begin{minipage}[t]{0.325\textwidth}
    \includegraphics[width=\textwidth]{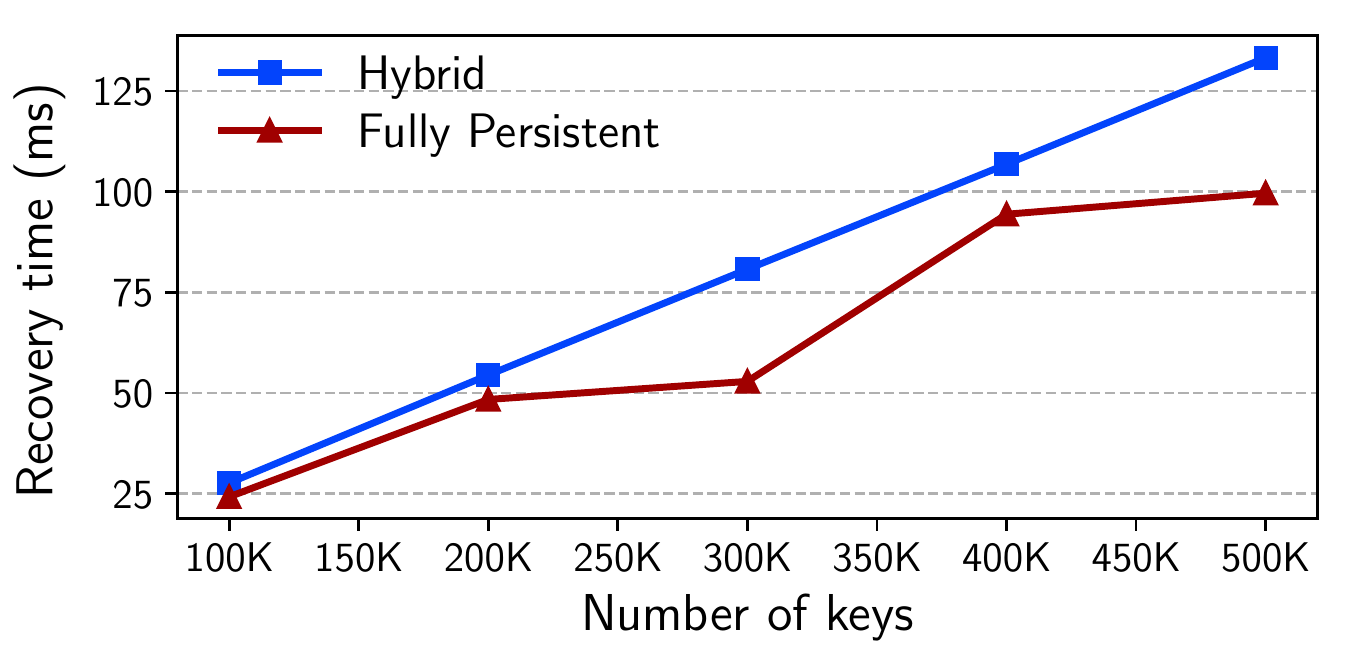}
    \vspace{-0.3in}
    \caption{Memcached recovery time.}
    \label{fig:memcached_recovery}
  \end{minipage}
  \vspace{-0.1in}
\end{figure*}

\vspace{0.05in}
\noindent{\textbf{Scalability}:} Figures~\ref{fig:redis_ycsb_client_throughput}
and \ref{fig:memcached_ycsb_client_throughput} show how the Redis and Memcached
designs scale as the YCSB client threads increase. We can see 
that our ported designs preserve the scalability characteristics of the
base systems. The Hybrid and Fully Persistent designs have similar
behavior when increasing the client threads. In Redis, they 
both hit the scalability bottleneck with 4 threads while the base Redis 
stops scaling at 6 threads. We suspect this is in part due to contention 
in the CPU. For Memcached, all three designs scale to 16 threads, 
demonstrating its multithreading advantage.

\vspace{-0.1in}
\subsection{Recovery Performance}
\vspace{-0.05in}
In this experiment, we insert a variable number of keys (100K to 10M), shutdown
and restart the system, and measure the time to recover the system as a function 
of the size of the key-value store. The recovery process revalidates pointers and 
reattches persisent memory allocations and rebuilds volatile data structures as 
necessary. The recovery finish point is when the system properly restores all the 
key-values it persisted before the shutdown or failure.

Figure \ref{fig:redis_recovery_time} shows that the recovery time of all variants
increases linearly in the size of the keyspace and that Hybrid Redis takes
around 20--25$\times$ as long to recover as does Fully Persistent Redis.
With 10 million keys, Hybrid Redis takes 28.5 seconds to recover all the
data, whereas Fully Persistent Redis only takes 4 seconds---a 7$\times$ difference.
Fully Persistent Redis recovers in a single pass over the memory space to rewrite pointers.
Hybrid Redis has to (1) iterate through the \textsf{PMDK} list of allocation 
pointer and (2) reinsert all keys into the hashtable to recover the pointers.
We break down the recovery time of Hybrid Redis by the two steps. 
Figure \ref{fig:redis_recovery_time} shows that the majority of the
recovery time comes from the \textsf{libpmemobj} iteration.



%

Recovery in Memcached shows different structure with Hybrid Memcached 
recovering nearly as fast as Fully Persistent Memcached (Figure~\ref{fig:memcached_recovery}). 
Rebuilding the hash table ends up being much faster as there are no new memory allocations.
The hash table pointers are updated in place in the persistent items.
Recovery of both variants is substantially slower than Fully Persistent Redis.
In Fully Persistent Memcached, there is more recovery work to do because there are more pointers
for Memcached more complex data structures.



We note that the differences in the design of Redis and Memcached may lead to different decisions
when choosing between Hybrid and Fully Persistent Designs.  Redis has much larger overheads for 
recovery in the Hybrid design to realize a comparable increase in operational throughput.

\vspace{-0.1in}
\subsection{Data Loss}
\vspace{-0.05in}
In order to ensure that our systems were crash consistent we used transactions over each
single key operation. This way we can make sure that no more than 1 key value pair 
is lost upon system failure. However, with volatile systems that rely on less 
consistent persistence options such as Base Redis' RDB feature, we saw a much 
larger data loss. We set Base Redis w/ RDB to snapshot its state to disk every 3 seconds.
We crash the system at set intervals of 2.5, 5, 10 and 15 seconds. With a transactional 
interface, Fully Persistent Redis and Hybrid Redis were able to show only 1 key value pair loss
while Base Redis with RDB suffered heavier losses: it lost 306,752 (all 
items), 797,239, 687,864, 853,219 key-value pairs respectively under the
four crash points.

\vspace{-0.1in}
\subsection{Porting Efforts}
\vspace{-0.05in}
\begin{table}[t]
  \footnotesize
  \centering
  \begin{tabular}{@{}lll@{}}
    \toprule
      System & Fully Persistent & Hybrid\\
    \midrule
      Memcached & 385 (6/55 files) & 371 SLOC (6/55 files) \\ 
      Redis & 727 (10/134 files) & 555 SLOC (10/134 files) \\
    \bottomrule
  \end{tabular}
  \caption{Modifications to the original systems.}
  \label{tab:porting_efforts}
\end{table}
Table~\ref{tab:porting_efforts} shows the modifications in lines of code we made 
to the base systems. We meet our goal in making our changes small and
less disruptive. We touched more lines of code with the Fully Persistent designs. 
This is in main part due to the sheer amount of data structures and corresponding 
functions that we had to convert to persistent memory. 

In terms of complexity, in developing Fully Persistent implementations, we 
have to be diligent in tracking data dependencies among the inter-related
data structures. With a large codebase, it is ease to miss making some dependent 
variables persistent and introduce partial inconsistency bugs. We had to
iteratively make the design correct.


The main porting challenge that came from the Hybrid Design was finding a way
of organizing persistent data across restart without an indexing structure. If
we had implemented this with an auxiliary data structure we would have
experienced much more SLOC as maintaining such a structure is not simple.
However, by leveraging pmdk's allocator linked list, we were able to reduce the
amount of effort.

\vspace{-0.1in}
\begin{table}[t]
  \setlength{\tabcolsep}{5pt}
  \footnotesize
  \centering
  \begin{tabular}{@{}llll@{}}
    \toprule
     Fully Persist. Redis & Hybrid Redis & PMEM-Redis & libpmemlog-AOF \\
    \midrule
      \SI{68828}{op/s} & \SI{126510}{op/s} & \SI{233744}{op/s} & \SI{144466}{op/s}\\ 
    \bottomrule
  \end{tabular}
  \begin{tabular}{@{}lll@{}}
    \toprule
     Fully Persist. Memcached & Hybrid Memcached & Lenovo Memcached \\
    \midrule
      \SI{56747}{op/s} & \SI{82826}{op/s} & \SI{66546}{op/s}\\ 
    \bottomrule
  \end{tabular}
  \caption{Average Throughputs of our ported designs and existing ports.}
  \label{tab:throughput_comparison_related}
\end{table}

\subsection{Comparison with Other PMEM Redis and Memcached}
We compare our ported designs with several open source PM ports of Redis
and Memcached. Table~\ref{tab:throughput_comparison_related} shows
the average throughputs. 
PMEM-Redis~\cite{PMEMRedis} writes values to persistent memory that are larger 
than an NVM threshold size while keeping smaller values in volatile memory 
(Section~\ref{sec:related_work}). Not surprisingly, this design is 
faster than both of our Redis implementations. Its better performance 
comes at a cost of severe data loss, whereas our implementations lose
at most 1 key-value pair.
Similarly, libpmemlog-AOF~\cite{libpmemlog-AOF} 
uses a persistent AOF for recovery allowing its system to be
slightly faster than our Hybrid Redis implementation. From a recovery
standpoint, our Fully Persistent Redis outperformed libpmemlog-AOF and
PMEM-Redis by over 3$\times$ while our Hybrid Redis was around 40-50\% slower.
Lenovo's Memcached implementation~\cite{LenovoMemcached} shows throughput 
17\% better than our Fully Persistent Memcached but 24.4\% slower than our 
Hybrid Memcached implementation due to their implementation choosing to persist 
entire items to their persistent slabs. 

Table~\ref{tab:porting_efforts} shows the modifications in the related work. 
Our Memcached modifications are smaller than Lenovo's. This is mainly because 
Lenovo's pmemcached is using the low-level \textsf{PMDK} interfaces such as 
\texttt{pmem\_flush} and \texttt{pmem\_persist}. To ensure atomicity, it
has to add additional sanity check fields into the persistent data 
structures,suck as checksums, validity bit and linked flag. Upon restart, it will examine
and discard potentially inconsistent data. We use the transaction 
interfaces of \textsf{PMDK}, which significantly simplify our 
modifications. We also notice that, due to the lack of failure-atomic 
transactions, even with the sanity checks, the Lenovo pmemcached can 
still incur partial inconsistencies when there is a untimely crash: e.g., 
the \texttt{time} field of an item.

\begin{table}[t]
  \footnotesize
  \centering
  \begin{tabular}{@{}lll@{}}
    \toprule
      Work & Modifications \\
    \midrule
      PMEM-Redis~\cite{PMEMRedis} & 996 (24/123 files) \\ 
      libpmemlog-AOF~\cite{libpmemlog-AOF} & 301 (5/118 files) \\
      WHISPER-redis$^*$~\cite{SnalliRedis} & 409 (10/97 files) \\ 
      Lenovo-pmemcached~\cite{LenovoMemcached} & 859 (8/52 files) \\
    \bottomrule
\end{tabular}
\caption{Porting efforts in other related work. *: could not recover properly upon Redis restart.}
  \label{tab:porting_efforts}
\end{table}

\section{Discussions}
Through the implementation of Hybrid and Fully Persistent versions of Redis and
Memcached, we summarize three principles for porting volatile KV stores to 
persistent memory. 

\emph{A hybrid design is preferable}. Although keeping all relevant indexing structures
  persistent greatly speeds up the recovery, the Fully Persistent design suffers
  from significant performance overhead. For many modern KV stores that receive 
  a large amount of requests, ensuring large operational throughput, quick turnaround 
  time, and good tail latency is of utmost importance to users. Even though
  hybrid designs recover slower, its absolute recovery time is still compelling 
  (for 10M keys, Hybrid Redis can recover in 28 seconds).
  
\emph{Persistent data structures should be allocated in large chunks to amortize the 
  increased latency of persistent memory}. One of the major differences between Redis and
  Memcached that heavily influenced the porting procedure to persistent memory
  was the differing allocation schemes (per-key/value-pair vs. slab allocation). 
  Just as a designer should aim to reduce the number of writes in persistent memory, 
  they should also aim to reduce the number of allocations in persistent memory due 
  to its high cost and performance inefficiency. Having a per-key/value pair allocation 
  scheme in a volatile KV store is still reasonable as the performance costs 
  of volatile allocation are negligible and the complexity of managing larger allocations can be
  cumbersome. However, persistent allocations require writes to PM, which incurs a 
  much larger performance cost than volatile writes.

\emph{Full featured persistent memory libraries ease development and lead to simple 
    implementations.} In order to have a method of reading persistent data upon 
    restart, developers either have to maintain their own auxiliary persistent 
    data structure or rely on their persistent memory library to recollect data. 
    While some libraries might store persistent data 
    contiguously and make any reads from persistent files inconsequential, other libraries such as \textsf{PMDK} may 
    require one to make some modifications to keep track of persistent data addresses.
    Without relying on a persistent memory library, developers will have to
    manually create their own persistent data storage method or structure.
    These findings are directly taken from our observations when we  maintained
    our own auxiliary data structure for Hybrid Redis. We found that creating
    our own data structure was not only a significant development effort to
    maintain but was also difficult to keep efficient.  As a result, we
    switched to using the \textsf{PMDK} allocator iterator for Hybrid Redis. 


\section{Conclusion}
With the combination of our empirical evaluation and guiding principles, we
were able to show that the hybrid design encapsulated the operational
performance needs of storage systems while the fully persistent design
optimizes the recovery performance of storage systems.  Hybrid Redis 
demonstrated 2$\times$ better operational throughput and 4$\times$ better tail latency 
compared to Fully Persistent Redis. However, Fully Persistent Redis
can recover 10 million keys 7$\times$ faster compared to Hybrid Redis.
Hybrid Memcached had 1.45$\times$ better operational throughput and 7$\times$
better tail latency compared to Fully Persistent Memcached. Fully Persistent 
Memcached had 33\% faster recovery than Hybrid Memcached.
We also gathered an additional 3 actionable design principles to keep in mind
when porting persistent KV Stores that carried over various systems. 
We conclude that when porting legacy systems to persistent memory, developers should
consider the hybrid design, a combination of volatile and nonvolatile data
structures, when prioritizing operational performance and the fully persistent
design, keeping all data structures in nonvolatile memory, for recovery
purposes.

{
\bibliographystyle{abbrv-customized}
\bibliography{reference}
}

\end{document}